\documentclass[eqsecnum,showpacs,pre,twocolumn]{revtex4}%
\usepackage{graphicx}
\usepackage{amsmath}
\usepackage{bm}
\usepackage{amsfonts}
\usepackage{amssymb}%
\begin{document}
\title{A Generalized Epidemic Process and Tricritical Dynamic Percolation}
\author{Hans-Karl Janssen}
\affiliation{Institut f\"{u}r Theoretische Physik III, Heinrich-Heine-Universit\"{a}t,
40225 D\"{u}sseldorf, Germany}
\author{Martin M\"uller}
\affiliation{Institut f\"{u}r Theoretische Physik III, Heinrich-Heine-Universit\"{a}t,
40225 D\"{u}sseldorf, Germany}
\author{Olaf Stenull}
\affiliation{Department of Physics and Astronomy, University of Pennsylvania, Philadelphia,
Pennsylvania 19104, USA}
\date{\today}

\begin{abstract}
The renowned general epidemic process describes the stochastic evolution of a
population of individuals which are either susceptible, infected or dead. A
second order phase transition belonging to the universality class of dynamic
isotropic percolation lies between endemic or pandemic behavior of the
process. We generalize the general epidemic process by introducing a fourth
kind of individuals, viz.\ individuals which are weakened by the process but
not yet infected. This sensibilization gives rise to a mechanism that introduces a global
instability in the spreading of the process and therefore opens the
possibility of a discontinuous transition in addition to the usual continuous
percolation transition. The tricritical point separating the lines of first and
second order transitions constitutes a new universality class, namely the universality class of tricritical dynamic isotropic percolation. Using renormalized field theory we work out a detailed
scaling description of this universality class. We calculate the scaling exponents in an
$\varepsilon$-expansion below the upper critical dimension $d_{c}=5$ for
various observables describing tricritical percolation clusters and their
spreading properties. In a remarkable contrast to the usual percolation
transition, the exponents $\beta$ and ${\beta}^{\prime}$ governing the two
order parameters, viz.\ the mean density and the percolation probability,
turn out to be different at the tricritical point. In addition to the scaling
exponents we calculate for all our static and dynamic observables logarithmic
corrections to the mean-field scaling behavior at $d_c=5$.
\end{abstract}
\pacs{64.60.Ak, 05.40.-a, 64.60.Ht, 64.60.Kw}
\maketitle

\section{Introduction}

\label{sec:intro}
The formation and the properties of random structures have
been a major theme in statistical physics for many years. In case the
formation of such structures obeys local rules, these processes can often be
expressed in the language of population growth. It is well known that two
particular growth processes lead to random structures with the properties of
percolation clusters. The so called simple epidemic process (SEP) leads to
directed percolation (DP)~\cite{BrHa57,CaSu80,Ob80,Hin00}. The SEP is also
known as epidemic with recovery, Gribov process~\cite{GraSu78,GraTo79}, as the
stochastic version of Schl\"{o}gls first reaction~\cite{Ja81,Schl72} or in
elementary particle physics as Reggeon field theory~\cite{Gri67,GriMi68,Mo78}.
The so called general epidemic process (GEP)~\cite{Mol77,Bai75,Mur89}, also
known as epidemic with removal, generates isotropic percolation (IP)
clusters~\cite{Gra83,Ja85,CaGra85,StAh92BuHa96} and models therefore the
universality class of dynamic isotropic percolation (dIP).

Epidemic models like the SEP and the GEP are relevant for a wide range of
systems in physics, chemistry, biology, and sociology. Undoubtedly, the
potential of such simple models has its limitations because they rely on
strong simplifying assumptions such as the homogeneity of the substrate~\cite{footnote_quenchedDisorder},
isotropy of the infections, immobility of individuals and so on. However, the
transition between population survival and extinction of these processes is a
nonequilibrium continuous (second order) phase transition phenomenon and hence
is characterized by \emph{universal} scaling laws which are shared by entire
classes of systems. Near these transitions, simplistic epidemic models like
the SEP and the GEP are of great value, because they are powerful workhorses
to study the mutual properties of their entire universality class which also
should include more realistic models.

The universal properties of DP and dIP are well known today thanks to numerous studies of the SEP and GEP, respectively. Relatively little is known, on the other hand, whether and under what modifications these stochastic growth processes allow for first order phase transitions between
their endemic and pandemic states, and by the same token, for tricritical behavior at the phase-space boundary between first and second order transition. In the context of DP, these questions where addressed to some extend by Ohtsuki and Keyes~\cite {OhKe87}. In this paper we will study this intriguing topic in the context of dIP by generalizing the famous GEP.

The standard GEP, assumed for simplicity to take place on a lattice
\cite{Gra83}, can be described with help of the reaction scheme
\begin{subequations}
\label{GEP}%
\begin{align}
S(\mathbf{n})+X(\mathbf{m})\quad &  \overset{\kappa}{\longrightarrow}\quad
X(\mathbf{n})+X(\mathbf{m})\, ,\label{reactA}\\
X(\mathbf{n})\quad &  \overset{\lambda}{\longrightarrow}\quad E(\mathbf{n})\,
, \label{reactB}%
\end{align}
with reaction rates $\kappa$ and $\lambda$. $S$, $X$, and $E$ respectively
denote susceptible, ill, and dead (or immune) individuals on nearest neighbor
sites $\mathbf{n}$ and $\mathbf{m}$. A susceptible individual may be infected
by an ill neighbor with probability $\kappa$ [reaction~(\ref{reactA})]. By
this mechanism the disease (henceforth also called the agent) spreads
diffusively. Ill individuals die with a probability $\lambda$
[reaction~(\ref{reactB})]. There is no healing of infected individuals and no
spontaneous infection. In a finite system the manifold of states without any
infected individual is inevitably absorbing. Whether a single infected site
leads to an everlasting epidemic in an infinite system depends on the ratio
$\kappa/\lambda$. With $\lambda$ fixed there is a certain value $\kappa
=\kappa_{c}$ that for all $\kappa>\kappa_{c}$ an eternal epidemic (a pandemic)
occurs. The probability $P(\kappa)$ for the occurrence of a pandemic as a
function of $\kappa$ goes to zero continuously at the critical point
$\kappa_{c}$. The behavior of the model near this critical point is in the
universality class of dIP.

Of course, one can conceive many alterations of the GEP. Some modifications will lead to models which still belong to the dIP universality class whereas other alterations will produce models belonging to other universality classes. As an example for the latter, one might think of mobile susceptible or immune individuals which, say, diffuse through space. The resulting models are reaction-diffusion type models which have nothing to do with dIP~\cite{footnote_diff_sick}. We are not interested in this kind of alterations. Rather, we are interested in modifications of the GEP that preserve the dIP universality class and which, for certain parameter values, allow for tricriticality and first order percolation transitions. To be more specific, we are interested in the most relevant mechanism leading to tricritical and first order dIP. In spirit our model is closely related to the canonical model for tricriticality in eqilibrium systems, viz.\ the $\phi^4$ model with an additional $\phi^6$ term where the free energy density is of the form $f= \tau \phi^2+ g_4 \phi^4 + g_6 \phi^6$ ~\cite{tomsBook}. If $g_4$ is positive then higher order terms including the $\phi^6$ term are irrelevant and one has in mean field theory a second order transition when $\tau$ passes through zero. Otherwise, however, one has an instability and higher order terms are required for stabilization of which the $\phi^6$ term is the most relevant one. For $g_4 < 0$ one has a first order transition at a critical value of $\tau$ that depends on $g_4$ and $g_6$ and the point $\tau=g_4=0$ constitutes a tricritical point. Our model to be defined in the next paragraph is so that it introduces a similar instability in the GEP and via this instability it allows for tricriticality and first order percolation. In its field theoretic description our model features, compared to the standard GEP, an additional higher order term which becomes the most relevant stabilizing term when the usual GEP coupling vanishes.

Our modification of the GEP can be defined in simple terms. The basic idea is to enrich the reaction scheme~(\ref{GEP}) by introducing \emph{weak} individuals $Y$. Instead of being infected right
away by an ill neighbor, any susceptible individual may be weakened with a
reaction rate $\mu$ by such an encounter. When the disease passes by anew, a
weakened individual is more sensitive and gets sick with a rate $\nu> \kappa$.
In the following we refer to this model as the generalized GEP (GGEP).
In addition to the reactions~(\ref{GEP}), the GGEP is described by the reactions
\end{subequations}
\begin{subequations}
\label{GGep}%
\begin{align}
S(\mathbf{n})+X(\mathbf{m})\quad &  \overset{\mu}{\longrightarrow}\quad
Y(\mathbf{n} )+X(\mathbf{m})\, ,\\
Y(\mathbf{n})+X(\mathbf{m})\quad &  \overset{\nu}{\longrightarrow}\quad
X(\mathbf{n} )+X(\mathbf{m})\, .
\end{align}
As we go along, we will show that the occurrence of the weak individuals gives
rise to an instability that can lead to a discontinuous transition and compact
(Eden \cite{Ed81,Ca83}) growth of the epidemic if $\nu$ is greater than a
critical value $\nu_{c}(\kappa,\mu)$. In the enlarged three-dimensional phase
space spanned by $\kappa$, $\mu$, and $\nu$ with fixed $\lambda$, there exists
a critical surface associated with the usual continuous percolation transition
and a surface of first order transitions characterized by a finite jump in the
probability $P(\kappa,\mu,\nu)$ for the occurrence of a pandemic. These two
surfaces of phase transitions meet at a line of tricritical points.

The focus of this paper lies on the universal properties of the GGEP near this
tricritical line. By using the methods of renormalized field theory we work
out a scaling description of the new universality class of tricritical dynamic
isotropic percolation (TdIP) to which the tricritical GGEP belongs. We study a
multitude of static and dynamic observables that play important roles in
percolation theory. In particular we calculate the critical exponents
describing the scaling behavior of these observables below 5 dimensions as well as logarithmic corrections to the mean-field scaling behavior in 5 dimensions.

The outline of our paper is as follows. In Sec.~\ref{meanFieldSec} we consider
the GGEP in a mean-field theory. As the main result of Sec.~\ref{meanFieldSec}%
, the mean-field analysis will reveal the structure of the phase diagram. With
the aim of studying the effects of fluctuations, we condense the principles
defining TdIP into a field theoretic model in Sec.~\ref{sec:FieldTheory}. In
Sec.~\ref{staticGGEP} we work out the scaling properties of static aspects of
TdIP. Section~\ref{dynGGEP} treats the dynamic scaling properties. Concluding remarks are provided in Sec.~\ref{sec:conclusion}. There is
one Appendix in which we sketch the calculation of a parameter integral that
is helpful in computing Feynman diagrams.

\section{Mean-Field Theory}
\label{meanFieldSec}

A mean-field description of the GGEP can be formulated by treating the reaction
equations (\ref{GEP}) and (\ref{GGep}) as deterministic equations without
fluctuations. This deterministic approximation leads to the system of
differential equations
\end{subequations}
\begin{subequations}
\label{meanfield}%
\begin{align}
\dot{S}(\mathbf{n},t)  &  =-(\kappa+\mu)S(\mathbf{n},t)\sum_{\mathbf{m}%
}^{nn(\mathbf{n} )}X(\mathbf{m},t)\, ,\label{meanS}\\
\dot{Y}(\mathbf{n},t)  &  =\Big(\mu S(\mathbf{n},t)-\nu Y(\mathbf{n}%
,t)\Big)\sum_{\mathbf{m }}^{nn(\mathbf{n})}X(\mathbf{m},t)\, ,\label{meanY}\\
\dot{X}(\mathbf{n},t)  &  =\Big(\kappa S(\mathbf{n},t)+\nu Y(\mathbf{n}%
,t)\Big)\sum_{ \mathbf{m}}^{nn(\mathbf{n})}X(\mathbf{m},t)\nonumber\\
&  -\lambda X(\mathbf{n},t)\, ,\label{meanX}\\
\dot{E}(\mathbf{n},t)  &  =\lambda X(\mathbf{n},t) \label{meanE}%
\end{align}
governing the dynamics of the different kinds of individuals. Here,
$\sum_{\mathbf{m}}^{nn(\mathbf{n})}$ denotes summation over the nearest
neighbors of $\mathbf{n}$. At each lattice site there is the additional
constraint
\end{subequations}
\begin{align}
\label{constraint}S+X+Y+E=1\, .
\end{align}
Thus, $S$, $X$, $Y$, and $E$ can be interpreted as the probability of finding
the corresponding state at a site $\mathbf{n}$. Note that the processes can
only proceed at places where the probability to find ill individuals in the
neighborhood is not zero. We use the canonical initial condition that all
sites of the initial state are susceptible except for the site at the origin
which is ill, $X_{0}(\mathbf{n})=\delta_{\mathbf{n},0}$ and $Y_{0}%
(\mathbf{n})=E_{0}(\mathbf{n})\equiv0$.

Equations~(\ref{meanS}) and (\ref{meanY}) are are readily integrated. We
obtain
\begin{equation}
S(\mathbf{n},t)=S_{0}(\mathbf{n})\exp\bigg(-\rho\sum_{\mathbf{m}%
}^{nn(\mathbf{n})}E( \mathbf{m},t)\bigg)
\end{equation}
and
\begin{align}
\mu S(\mathbf{n},t)+(\rho-\nu)Y(\mathbf{n},t)  &  =\Big(\mu S_{0}%
(\mathbf{n})+(\rho-\nu)Y_{0}(\mathbf{n})\Big)\nonumber\\
&  \times\exp\bigg(-\nu\sum_{\mathbf{m}}^{nn(\mathbf{n})}E(\mathbf{m},t)
\bigg)\, , \label{simplesol}%
\end{align}
where we have defined $\rho=\kappa+\mu$. The time scale $\lambda$ has been set
to unity for simplicity. Equation~(\ref{meanE}) together with the
constraint~(\ref{constraint}) leads finally to the mean-field equation of motion of
the GGEP,
\begin{align}
\dot{E}(\mathbf{n},t)  &  =1-E(\mathbf{n},t)\nonumber\\
&  -\bigg\{ \frac{\rho-\kappa}{ \rho-\nu} \,\exp\bigg(-\nu\sum_{\mathbf{m}%
}^{nn(\mathbf{n})}E(\mathbf{m},t)\bigg)\nonumber\\
&  +\frac{\kappa-\nu}{\rho-\nu}\exp\bigg(-\rho\sum_{\mathbf{m}}^{nn(\mathbf{n}
)}E(\mathbf{m},t)\bigg)\bigg\} \, S_{0}(\mathbf{n})\, . \label{GGEPstate}%
\end{align}
In the asymptotic regime $|\mathbf{n}|,t\rightarrow\infty$ one can neglect time
and space dependence in (\ref{GGEPstate}) and use the approximation $\sum_{
\mathbf{m}}^{nn(\mathbf{n})}E(\mathbf{m},t)\approx zE(\mathbf{n},t)$, $z$
being the coordination number of the lattice. Hence the asymptotic values of
$E$ are the solutions of the equation of state
\begin{equation}
E=\frac{\rho-\kappa}{\rho-\nu}\left(  1-\mathrm{e}^{-z\nu E}\right)  +\frac{
\kappa-\nu}{\rho-\nu}\left(  1-\mathrm{e}^{-z\rho E}\right)  =:f(E)\, .
\label{asymptsol}%
\end{equation}
By setting $\rho=\kappa$, corresponding to $\mu=0$, one obtains the equation
of state of the usual GEP \cite{Gra83}
\begin{equation}
E=1-\mathrm{e}^{-z\kappa E} \label{GEPstate}%
\end{equation}
with $\kappa=\kappa_{c}=1/z$ determining the second order phase transition
corresponding to ordinary isotropic percolation. For $\kappa<\kappa_{c}$,
Eq.~(\ref{GEPstate}) has only the solution $E=0$ which means that the disease
does not percolate, i.e., is endemic. In the other case, $\kappa>\kappa_{c}$, a stable solution
$E>0$ arises, signalling the percolative pandemic character of the disease.
These types of solutions exist also in the full equation of
state~(\ref{asymptsol}). Using $\kappa\leq\rho$, one demonstrates easily that
$f(E)$ increases monotonically from zero to one in the interval $0\leq
E<\infty$. Thus, $E=0$ is always a solution of Eq.~(\ref{asymptsol}). It
follows from the equation of motion~(\ref{GGEPstate}) that only solutions of
Eq.~(\ref{asymptsol}) with $f^{\prime}(E)<1$ are stable. Because $f^{\prime
}(0)=z\kappa$, a stable percolating solution $E>0$ exists always for
$\kappa>\kappa_{c}$. The existence of more than one nontrivial solution
requires necessarily that $f^{\prime\prime}(E)=0$ at least for one value
$E>0$. From Eq.~(\ref{asymptsol}) we derive the inequality
\begin{equation}
f^{\prime\prime}(E)/z^{2}\leq\Big(\rho\nu-\kappa(\rho+\nu)\Big)\exp
(-zE\max(\rho,\nu))\, .
\end{equation}
With $f^{\prime\prime}(E)<0$ for $E\gg1$ and $f^{\prime\prime}(0)/z^{2}%
=\rho\nu-\kappa(\rho+\nu)$ we therefore find
\begin{equation}
\frac{1}{\kappa}>\frac{1}{\rho}+\frac{1}{\nu}\ ,\quad\kappa<\kappa_{c}
\label{unstable}%
\end{equation}
as the necessary and sufficient conditions for the existence of a second
locally stable solution $E>0$ besides $E=0$ with a discontinuous transition
between the two. The line of tricritical points where the first and the second
order transitions meet in the $\kappa$-$\rho$-$\nu$ phase space is determined
by $1/\kappa=1/\rho+1/\nu=z$.

In the following we focus on the phenomena arising near the tricritical line.
In this region of the phase space, $E(\mathbf{n,}t)$ is, except for a microscopic region around the origin $\mathbf{n}=\mathbf{0}$ of the position space, small and slowly varying. Hence, we may approximate Eq.~(\ref{GGEPstate}) by the deterministic reaction-diffusion
equation
\begin{align}
\dot{E}(\mathbf{x},t)  &  =D\nabla^{2}E(\mathbf{x},t)-\lambda\bigg[\tau
E(\mathbf{x},t)\nonumber\\
&  + \frac{f}{2}E(\mathbf{x},t)^{2}+\frac{g}{6}E(\mathbf{x},t)^{3}\bigg] ,
\label{GGEPdiffusion}%
\end{align}
where $D=\kappa a^{2}$, with $a$ being the lattice constant, $\tau
=1-\kappa/\kappa_{c}$, $f=\kappa\rho\nu(1/\rho+1/\nu-1/\kappa)/\kappa_{c}^{2}$, and $g=(\rho\nu(\rho+\nu+\kappa)-\kappa(\rho+\nu)^{2})/\kappa_{c}^{3}\approx\rho\nu/\kappa_{c}^{2}$. As a consequence of Eq.~(\ref{meanE}) we have the constraint $\dot{E}(\mathbf{x},t)\geq0$. 

\begin{figure}[ptb]
\includegraphics[width=7.5cm]{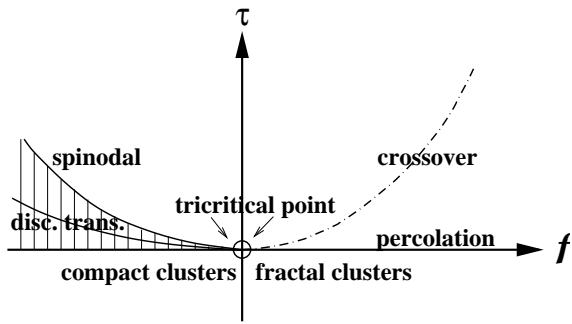}
\caption{The mean-field phase diagram.}
\label{mfDia}
\end{figure}

Holding the positive coupling $g$ constant, Eq.~(\ref{GGEPdiffusion}) comprises only two tunable parameters, viz.\ $\tau$ and $f$ and hence the dimensionality of the phase diagram is reduced to two. As follows from the different types of solutions of Eq.~(\ref{GGEPdiffusion}) to be discussed in a moment, this two-dimensional phase diagram features a line of second
order transitions ($\lambda$-line), a line of first order transitions and a
tricritical point determined by $\tau=f=0$ separating the two lines of transitions. In addition, there are 2 spinodal lines. The entire phase diagram is depicted in Fig.~\ref{mfDia}.

Equation~(\ref{GGEPdiffusion}) has besides the trivial solution $E=0$, which
is stable for $\tau>0$, a nontrivial locally stable stationary homogeneous
solution $E=A$ with
\begin{equation}
A=\Big(\sqrt{9f^{2}/4-6g\tau}-3f/2\Big)\, , \label{Ampl}%
\end{equation}
which is physical only if $A>0$. For $f>0$ one has therefore a continuous
transition from $E=0\ (\tau>0)$ to $E=A\approx2\left\vert \tau\right\vert /f$
$(\tau<0)$. However, for $f<0$ there is a certain value $\tau=\tau_{d} $
between the spinodals $\tau=3f^{2}/8g$ and $\tau=0$ for the discontinuous
transition from $E=0$ to $E=A$. The values of $\tau_{d}$ may be determined by
studying a travelling wave solution of the equation of motion
(\ref{GGEPdiffusion}) that describes the infection front of a big expanding
sherical cluster of the epidemic. Such a solution is given by
\begin{subequations}
\label{travwav}%
\begin{align}
E(x,t)  &  =A\Big(1-\tanh(b(x-vt))\Big)/2\ ,\\
X(x,t)  &  =\dot{E}(x,t)=\frac{vbA}{2\bigl[\cosh(b(x-vt))\bigr]^{2}}%
\end{align}
with
\end{subequations}
\begin{equation}
b=\sqrt{\frac{\lambda g}{48D}}\,A\,,\quad v=\sqrt{\frac{3\lambda D}{4g}%
}\Big(\sqrt{9f^{2}/4-6g\tau}+f/2\Big)\ . \label{travwav2}%
\end{equation}
The condition $X(x,t)\geq0$ requires $v\geq0$. The first order transition from
$E=0$ to $E=A$ at $\tau_{d}$ is therefore defined by the phase equilibrium
condition $v=0$ leading to
\begin{equation}
\tau_{d}=\frac{f^{2}}{3g}\,. \label{transtemp}%
\end{equation}
$E$ jumps at $\tau=\tau_{d}$ from zero to the value $2\left\vert f\right\vert/g$.

\section{Field Theoretic Model}

\label{sec:FieldTheory}

In this section we will derive a dynamic response functional
\cite{Ja76,DeDo76,Ja92} for the GGEP based on very general arguments alluding
to the universal properties of TdIP. First we will distill the basic principles of percolation processes allowing for tricritical behavior. Next, we cast these principles into the form of a Langevin
equation. Then we refine the Langevin equation into a minimal field theoretic model.

As an alternative avenue to a field theoretic model for the GGEP one migtht be tempted to use the so-called ``exact'' approach which, as a first step, consists of reformulating the microscopic master-equation for the reactions~(\ref{GEP}) and (\ref{GGep}) as a bosonic field theory on the lattice.
The next and pivotal step in this approach is to take the continuum limit. Albeit the ``exact'' approach  with a naive continuum limit leads to a consistent dynamic functional for the GGEP (after deleting several irrelevant terms) this approach must be cautioned against. Strictly speaking, one has to use Wilson's statistical continuum limit \cite{Wils75} in the renormalization theory of critical phenomena. This procedure consists of successive coarse graining of the mesoscopic slow variables (order parameters and conserved quantities as functions of microscopic degrees of freedom), the elimination of fluctuating residual microscopic degrees of freedom, and a rescaling of space
and time. In general, microscopic variables do {\em not} qualify as order parameters. Alarming examples are the pair contact processes (PCP and PCPD), where a naive continuum limit of the microscopic master-equations leads to untenable critical models. Therefore, we devise our field theoretic model representing the TdIP universality class using a purely mesoscopic stochastic formulation based on the correct order parameters identified through physical insight in the nature of the critical phenomenon. Hence, our dynamic response functional stays in full analogy to the Landau-Ginzburg-Wilson functional and provides a reliable starting point of the field theoretic method.

\subsection{Langevin equation}

The essence of isotropic percolation processes can be summarized by four
statements describing the universal features of the evolution of such
processes on a homogeneous substrate. Denoting the density of the agents (the
infected individuals) by $n(\mathbf{r},t)$ and the density of the debris (the
immune or dead individuals) which is proportional to the density of the weakened substrate by $m(\mathbf{r},t)$, these four statements read:

\begin{enumerate}
\item[(i)] There is a manifold of absorbing states with $n\equiv0$ and corresponding
distributions of $m$ depending on the history of $n$. These states are
equivalent to the extinction of the epidemic.

\item[(ii)] The substrate becomes activated (infected) depending on the
density of the agents \emph{and} the density of the debris. This mechanism
introduces memory into the process. The debris ultimately stops the disease
locally. However, it is possible that the activation is strengthened by the debris through some mechanism (sensibilization of the substrate).

\item[(iii)] The process (the disease) spreads out diffusively. The agents
(the activated substrate) become deactivated (converted into debris) after a short time.

\item[(iv)] There are no other slow variables. Microscopic degrees of freedom can be summarized into a local noise or Langevin force $\zeta(\mathbf{r},t)$ respecting the first statement (i.e., the noise cannot generate agents).
\end{enumerate}

The general form of a Langevin equation resembling these statements is given
by
\begin{subequations}
\label{eq:General Langevin}%
\begin{align}
\lambda^{-1}\dot{n}  &  =\nabla^{2}n+R(n,m)\,n+\zeta
\,,\label{eq:General Langevin-1}\\
m(\mathbf{r},t)  &  =\lambda\int_{-\infty}^{t}n(\mathbf{r},t^{\prime
})\,dt^{\prime}\,,
\end{align}
where $\lambda$ is a kinetic coefficient and the Gaussian noise correlation
reads
\end{subequations}
\begin{align}
\overline{\zeta(\mathbf{r},t)\zeta(\mathbf{r}^{\prime},t^{\prime})}  &
=\lambda^{-1}Q\big(n,m\big)\,n(\mathbf{r},t)\,\delta(\mathbf{r}-\mathbf{r}%
^{\prime})\,\delta(t-t^{\prime})\nonumber\label{eq:GeneralNoise}\\
&  -\lambda^{-1} \alpha n(\mathbf{r},t)\,\nabla^{2}\delta(\mathbf{r}%
-\mathbf{r}^{\prime})\,\delta(t-t^{\prime})\nonumber\\
&  +Q^{\prime}\big(n,m\big)n(\mathbf{r},t)n(\mathbf{r},t^{\prime}%
)\,\delta(\mathbf{r}-\mathbf{r}^{\prime})+\cdots\,.
\end{align}
The first row in Eq.~(\ref{eq:GeneralNoise}) represents time-local reaction noise. The second row describes noise originating from diffusion and the last row 
shows an example of possible time-non-local noise (quenched noise) that may be
acquired through random disorder or through the elimination of microscopic slow
variables, e.g., fluctuations of the debris. The structure of the 3 terms is so
that they respect the absorbing state condition. Of course many further contributions to Eq.~(\ref{eq:GeneralNoise}) are conceivable (hence the ellipsis in the third row)
including non-Markovian and also non-Gaussian noise. We will see below, that all these terms turn out being irrelevant, and that only the simplest form of reaction noise contributes the minimal field theoretic model. 

The dependence of the rate $R(n,m)$ on the density of the debris $m(\mathbf{r},t)$
describes memory of the process mentioned above. We are interested primarily in the behavior of the process close to the tricritical point, where $n$ and $m$ are small allowing for polynomial
expansions $R(n,m)=-\tau-an-fm-gm^{2}/2+\ldots$, $Q(n,m)=\gamma+\cdots$, and
$Q^{\prime}(n,m)=\gamma^{\prime}+\cdots$. The justification for the truncation
of the expansions will be given later by IR relevance-irrelevance arguments.
As long as $f>0$, the second order term $gm^{2}$ of the rate $R$ is irrelevant
near the transition point and the process models ordinary dynamic isotropic
percolation~\cite{Ja85,CaGra85}. We permit both signs of $f$ so that our model
accounts for sensibilization (weakened substrate) and allows for compact
(Eden) spreading. Consequently we need the second order term for stabilization
purposes, i.e., to limit the density $n$ to finite values.

\subsection{Dynamic response functional}

In order to apply the renormalization group (RG) and field-theoretic
methods~\cite{Am84,ZiJu93}, it is convenient to use the path-integral
representation of the underlying stochastic process $n(\mathbf{r},t)$
\cite{Ja76,DeDo76,Ja92}. With the imaginary-valued response field denoted by
$\widetilde{n}(\mathbf{r},t)$, the generating functional of the Green
functions (connected response and correlation functions) takes the form
\begin{align}
\mathcal{W}\bigl[H,\widetilde{H}\bigr]  &  =\ln\int\mathcal{D}\bigl[\widetilde
{n},n\bigr]\exp\Big\{-\mathcal{J}\bigl[\widetilde{n},n\bigr]\nonumber\\
&  +\int d^{d}x\int dt\bigl(\widetilde{H}n+H\widetilde{n}\bigr)\Big\}\,.
\label{PathInt}%
\end{align}
The generating field $H(\mathbf{r},t)$ corresponds to an additive source term
for the agent in the equation of motion~(\ref{eq:General Langevin}).
Therefore, the response function defined by a functional derivative with
respect to $H(\mathbf{r},t)$ describes the influence of a seed of the agent at
$(\mathbf{r},t)$. The dynamic functional $\mathcal{J}\bigl[\widetilde
{n},n\bigr]$ and the functional measure $\mathcal{D}\bigl[\widetilde
{n},n\bigr]\propto\prod_{\mathbf{x},t}\bigl(d\widetilde{n}(\mathbf{r}%
,t)dn(\mathbf{r},t)/2\pi i\bigr)$ are defined using a prepoint (Ito)
discretization with respect to time \cite{Ja92}. The prepoint discretization
leads to the causality rule $\theta(t\leq0)=0$ in response functions. This
rule will play an important role in our diagrammatic perturbation calculation
because it forbids response propagator loops (see below). Note that the path
integrals are always calculated with the initial and final conditions
$n(\mathbf{r},-\infty) = \widetilde{n}(\mathbf{r},\infty)=0$.

The stochastic process defined by Eqs.~(\ref{eq:General Langevin}) and
(\ref{eq:GeneralNoise}) leads via the expansions of $R$, $Q$, and $Q^{\prime}$
to the preliminary dynamic response functional
\begin{align}
\hspace{-.2cm}\mathcal{J}^{\prime}  &  =\int d^{d}x\,\biggl\{\lambda\int
dt\,\Big[\tilde{n}\bigl(\lambda^{-1}\partial_{t}-\nabla^{2}+\tau+an+fm\nonumber
\\&  \quad+\frac{g}{2}m^{2}\bigr)n
-\frac{\gamma}{2}n\tilde{n}^{2}-\alpha n\bigl(\nabla\tilde
{n}\bigr)^{2}\Big]-\frac{\gamma^{\prime}}{2}\Big[\lambda\int dt\,\widetilde
{n}n\Big]^{2}\biggr\}\,. \label{Jfull}%
\end{align}
As we have remarked above, the term proportional to $g$ may be neglected only
if the coupling $f$ is positive definite. In this case Eq.~(\ref{Jfull})
reduces to the response functional of usual dynamic
percolation~\cite{Ja85,CaGra85}. As in all models with an absorbing state
transition, the functional $\mathcal{J}^{\prime}$ includes a redundant
variable which has to be removed before any application of
relevance-irrelevance arguments since it has no definite scaling dimension.
This redundant variable is connected with the rescaling transformation
\begin{subequations}
\label{rescaling}%
\begin{align}
n  &  \rightarrow b\,n\,,\quad\widetilde{n} \rightarrow b^{-1}\widetilde
{n}\,,\quad\alpha  \rightarrow b\alpha\,,\\
a & \rightarrow  b^{-1}a\,,\quad f \rightarrow b^{-1}f\,,\quad g \rightarrow b^{-2}g\,,\quad \gamma
\rightarrow b\,\gamma\, ,
\end{align}
which leaves $\mathcal{J}^{\prime}$ invariant. Exploiting this invariance, we we may set $\gamma
=1$ which fixes the  redundancy. Of course, this is justified only if $\gamma$ is a finite positive quantity in the region of interest of the phase diagram.

For the steps to follow, we need to know the naive dimensions of the
constituents of $\mathcal{J}^{\prime}$ after the removal of the redundant
variable. As usual, we introduce a convenient external length scale $\mu^{-1}$
so that $\mathbf{r}\sim\mu^{-1}$ and $\lambda t\sim\mu^{-2}$. Exploiting that
$\mathcal{J}^{\prime}$ has to be dimensionless, we readily find
\end{subequations}
\begin{subequations}
\label{eq:NaiveDim}%
\begin{align}
\widetilde{n}  &  \sim\mu^{2}\,,\quad n\sim\mu^{d-2}\,,\quad m\sim\mu
^{d-4}\,,\\
\tau &  \sim\mu^{2}\,,\quad f\sim\mu^{6-d}\,,\quad g\sim\mu^{2(5-d)}\,,\\
a & \sim\mu^{4-d}\,,\quad\alpha  \sim\mu^{-2}\,,\quad\gamma^{\prime}\sim\mu^{4-d}\,.
\end{align}
\end{subequations}
The highest dimension $d$ at which any of the finite and positive couplings becomes marginal (acquires a vanishing naive dimension) corresponds to the upper dimension $d_{c}$ of the theory.
This dimension separates trivial mean-field critical behavior for $d>d_{c}$ from non trivial
behavior in the regime where $d<d_{c}$ and where the relevant variable $\tau$
is small. Thus, if $f$ is finite and positive it follows that $d_{c}=6$. Then
$g$, $a$, $\gamma^{\prime}$, and $\alpha$ have negative naive dimensions and are
therefore irrelevant. The corresponding terms in the response functional
(\ref{Jfull}) vanish at the critical fixed point, and the response functional
displays the asymptotic symmetry $f^{-1/2}\widetilde{n}(\mathbf{r}%
,t)\leftrightarrow-f^{1/2}m(\mathbf{r},-t)$ \cite{Ja85}. The resulting
response functional is that of the GEP \cite{Ja85,CaGra85}.

However, if $f$ is zero as it is at the tricritical point, we must use
$g$ to fix the upper critical dimension which leads to $d_{c}=5$. The dimensions of
$a$, $\gamma^{\prime}$, and $\alpha$ are negative near $d_{c}$. Thus, diffusional
and quenched randomness of the noise is irrelevant here as they are for the GEP.
Dimensional analysis also justifies the truncation of the expansions of $P,$
$Q$, and $Q^{\prime}$, as well as the elimination of other terms. All higher
order terms are irrelevant in the IR renormalization group approach because
they carrying a \emph{negative\/} naive dimension near $d=d_{c}$. Collecting, we obtain the dynamic
response functional
\begin{align}
\mathcal{J}  &  =\int d^{d}x\,\lambda\int dt\,\tilde{n}\Big(\lambda^{-1}
\partial_{t}-\nabla^{2}+\tau\nonumber\\
&  \qquad\qquad+fm+\frac{g}{2}m^{2}-\frac{1}{2}\tilde{n}\Big)n \label{J}%
\end{align}
as our minimal field theoretic model for the TdIP universality class. We shall see as we move along, indicating the consistency of our model, that the proper elimination of IR irrelevant terms has led to an UV renormalizable theory at and below five dimensions.

Note that, contrary to dIP, the functional~(\ref{J}) does not have an asymptotic symmetry that relates
the two fields $\tilde{n}$ and $n$. Thus, these fields will get independent different anomalous dimensions leading to two different order parameter exponents.

Before we go on, we extract the diagrammatic elements implicit in
$\mathcal{J}$. These elements will play a central role later on when we
calculate the Green functions perturbatively. As usual, the Green functions
are the cumulants of the fields $n$ and $\widetilde{n}$ which correspond in graphical perturbation expansions to the sums of connected diagrams. Their actual calculations
are performed most economically in a time-momentum representation. To this end
we will use the spatial Fourier transforms of the fields $n$ and
$\widetilde{n}$ defined via
\begin{equation}
n(\mathbf{r},t)=\int_{\mathbf{q}}\mathrm{e}^{i\mathbf{q}\cdot\mathbf{r}%
}n_{\mathbf{q}}(t)\, ,\quad\widetilde{n}(\mathbf{r},t)=\int_{\mathbf{q}%
}\mathrm{e}^{i\mathbf{q}\cdot\mathbf{r}}\widetilde{n}_{\mathbf{q}}(t)\, ,
\label{eq:FourierTrans}%
\end{equation}
where $\int_{\mathbf{q}}\cdots:=(2\pi)^{-d}\int d^{d}q\cdots$. In the
time-momentum representation we can simply read off the diagrammatic elements
from the dynamic functional. The harmonic part of $\mathcal{J}$ comprises the
Gaussian propagator
\begin{subequations}
\label{eq:Propagator}%
\begin{align}
\langle n_{\mathbf{q}}(t)\widetilde{n}_{\mathbf{q}^{\prime}}(t^{\prime
})\rangle_{0}  &  =(2\pi)^{d}G(\mathbf{q},t-t^{\prime})\,\delta(\mathbf{q}%
+\mathbf{q}^{\prime})\,,\\
G(\mathbf{q},t)  &  =\theta(t-t^{\prime})\,\exp\Big(-\lambda\big(\tau
+\mathbf{q}^{2}\big)t\Big)\,,
\end{align}
where $\langle\cdots\rangle_{0}$ indicates averaging with respect to the
harmonic part of the dynamic functional~(\ref{Jfull}). The non-harmonic terms
give rise to the vertices $\lambda$, $-\lambda^{2}f\theta(t-t^{\prime})$ and
$-\lambda^{3}g\theta(t-t^{\prime})\theta(t-t^{\prime\prime})$. All four
diagrammatic elements of $\mathcal{J}$ are depicted in
Fig.~\ref{fullDynElements}.
\begin{figure}[ptb]
\includegraphics[width=7.5cm]{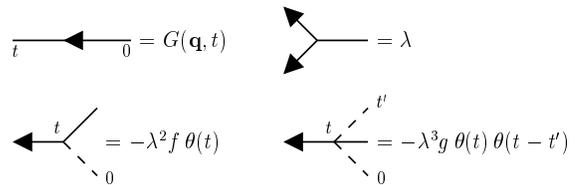}
\caption{The diagrammatic elements implied in the dynamic functional
$\mathcal{J}$.}%
\label{fullDynElements}
\end{figure}

\subsection{Quasi-static model}

As mentioned earlier, we are interested in the dynamic as well as the static,
i.e. $t\rightarrow\infty$, properties of tricritical percolation. Of course,
we could base our entire RG analysis on the full dynamic functional
$\mathcal{J}$ as given in Eq.~(\ref{Jfull}). Then we could extract the static behavior
from the dynamic behavior in the end by letting $t\rightarrow\infty$. This
would mean, however, that we had to determine all the required
renormalizations from dynamic Feynman diagrams composed of the
diagrammatic elements listed in Fig.~\ref{fullDynElements}. Fortunately, there
is a much more economic approach possible here which is based on taking the so-called quasi-static
limit. We will see shortly, that the perturbation theory
simplifies tremendously in this limit. All but one renormalization factors can
be calculated using this much simpler approach. Only for the one remaining
renormalization we have to resort to the dynamic response functional
$\mathcal{J}$. Taking the quasi-static limit amounts to switching the fundamental field
variable from the density of the agents to the density of the debris
$m(\mathbf{r}):=m(\mathbf{r},\infty)=\lambda\int_{-\infty}^{\infty
}dt\,n(\mathbf{r},t)$ left behind by the epidemic. Then, the static properties
of TdIP can be studied via the Green functions of the debris density. In particular the response functions $\langle\prod_{i}m(\mathbf{r}_{i})\widetilde{n}(0,0)\rangle$ and their connected counterparts  will be important for our analysis because they encode the static properties of the percolation cluster of the debris emanating from a seed localized at the origin at time zero.

After this prelude we now formally take the quasi-static limit of the dynamic
functional. The structure of $\mathcal{J}$ is so that we can directly let
\end{subequations}
\begin{equation}
\widetilde{n}(\mathbf{r},t)\rightarrow\widetilde{s}(\mathbf{r)\,,\quad}
m(\mathbf{r})=\lambda\int_{-\infty}^{\infty}dt\,n(\mathbf{r},t)\rightarrow
s(\mathbf{r})\,. \label{quaslim}%
\end{equation}
This procedure leads us from $\mathcal{J}$ to the quasi-static Hamiltonian
\begin{equation}
\mathcal{H}=\int d^{d}x\,\widetilde{s}\,\Big(\tau-\nabla^{2}+\frac{f}%
{2}s+\frac{g}{6}s^{2}-\frac{1}{2}\widetilde{s}\Big)\,s\,. \label{Hamilt}%
\end{equation}
It is easy to see that $\mathcal{H}$ generates each Feynman diagram that
contributes to $\langle\prod_{i}m(\mathbf{r}_{i})\prod_{j}\widetilde
{n}(\mathbf{\tilde{r}}_{j},0)\rangle$. Standing alone, however, this
Hamiltonian is not sufficient to describe the static properties of tricritical
isotropic percolation (TIP). As a remainder of its dynamical origin
$\mathcal{H}$ must be supplemented with the causality rule that forbids closed
propagator loops.

\begin{figure}[ptb]
\includegraphics[width=6.0cm]{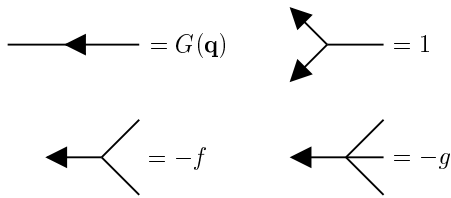}
%
\caption{The diagrammatic elements implied in the quasi-static Hamiltonian
$\mathcal{H}$.}%
\label{Elemente}
\end{figure}

The propagator of the quasi-static theory follows from Eq.~(\ref{Hamilt}) [or
likewise from Eq.~(\ref{eq:Propagator})] as
\begin{equation}
\langle s_{\mathbf{q}}\widetilde{s}_{\mathbf{q}^{\prime}}\rangle_{0}%
=(2\pi)^{d}G(\mathbf{q})\,\delta(\mathbf{q}+\mathbf{q}^{\prime})\, ,\quad
G(\mathbf{q})=\frac{1}{\tau+\mathbf{q}^{2}}\, . \label{statProp}%
\end{equation}
As far as vertices are concerned, $\mathcal{H}$ implies the three-leg vertices
$1$ and $-f$ and the four-leg-vertex $-g$. The quasi-static propagator and
vertices are shown in Fig.~\ref{Elemente}.
Knowing all the diagrammatic elements, one can straightforwardly check in
explicit graphical perturbation expansions that the quasi-static Green
functions calculated with $\mathcal{H}$ are equal diagram for diagram to the
zero-frequency limit of the dynamic Green functions (the Green functions of
the time-integrals of $n(\mathbf{r},t)$) calculated with $\mathcal{J}$.

Before embarking on our RG analysis, we finally mention the naive dimensions
of the quasi-static fields. These are given by
\begin{equation}
\widetilde{s}\sim\mu^{2}\,,\quad s\sim\mu^{d-4}\,. \label{NaiveFeldDim}%
\end{equation}

\section{Static Scaling Properties}

\label{staticGGEP}

Now we will study the static properties of TdIP, that is the properties of TIP. The dynamic properties of TdIP will be addressed later on.

\subsection{Diagrammatics}

In principle, we could extract the properties of TIP directly from the
correlation functions
\begin{align}
&  \langle s(\mathbf{r}_{1})\cdots s(\mathbf{r}_{N})\,\widetilde{s}%
(\mathbf{r}_{N+1})\cdots\widetilde{s}(\mathbf{r}_{N+\widetilde{N}}%
)\rangle\nonumber\\
&  =\int\mathcal{D}[s,\widetilde{s}]\,s(\mathbf{r}_{1})\cdots\widetilde
{s}(\mathbf{r}_{N+\widetilde{N}})\mathrm{e}^{-\mathcal{H}[\widetilde{s},s]}\,.
\label{StatGreen}%
\end{align}
However, since our model is translationally invariant, it is much more
convenient to use the vertex functions $\Gamma_{\widetilde{N},N}$ instead
which are related to the connected counterparts
\begin{align}
G_{N,\widetilde{N}}(\{\mathbf{r}\})=  &  \big\langle s(\mathbf{r}_{1})\cdots
s(\mathbf{r}_{N})\nonumber\\
&  \times\widetilde{s}(\mathbf{r}_{N+1})\cdots\widetilde{s}(\mathbf{r}%
_{N+\widetilde{N}})\big\rangle^{(\text{conn})} \label{GreenFu}%
\end{align}
of the correlation functions via Legendre transformation of their generating
functionals \cite{ZiJu93}. Graphically, the vertex functions consist of
amputated one-line irreducible diagrams. As usual in determining the
renormalizations, we can restrict ourselves to the superficially divergent
vertex functions, i.e., those vertex functions that have a non-negative $\mu
$-dimension at the upper critical dimension $d_{c}$. A simple dimensional
analysis shows that only the vertex functions that correspond to the different
terms of the Hamiltonian~(\ref{Hamilt}), viz.\ $\Gamma_{1,1}$, $\Gamma_{1,2}$,
$\Gamma_{2,1}$, and $\Gamma_{1,3}$, are primitively divergent. Thus, the
theory is renormalizable by additive and multiplicative renormalization of the
fields and the parameters of the theory.

Throughout, we will use dimensional regularization to calculate the Feynman
diagrams constituting the required vertex functions, i.e., we will compute the
diagrams in dimensions where they are finite for large momenta and then
continue the dimension analytically towards $d_{c}$. This procedure converts
the logarithmic large-momentum singularities of a cutoff regularization into
poles in the deviation $\varepsilon=d_{c}-d$ from $d_{c}$. However, polynomial large-momentum singularities that require additive renormalizations are unaccounted for in dimensional
regularization.  In order to discuss such additive
renormalizations, we will occasionally use a cutoff regularization
with a large momentum cutoff $\Lambda$. 

At this place it is worth stressing that real critical systems not involve any UV-divergencies because all
inverse wavelengths of fluctuations have a physical cutoff $\Lambda$. On the
other hand, critical systems suffer from IR-divergencies. However, if and only if we use a reliable field theory, we can formally transform  IR-divergencies into UV-divergencies by a simple rescaling. This way we can learn about IR-scaling properties of the critical system indirectly via the UV-renormalizations.  A correct and reliable statistical field theory constitutes what Wilson~\cite{Wils75} calls a logarithmic theory free of length scales. Only within such a theory it makes sense to apply the techniques of renormalized field theory to critical systems.

\subsubsection{Divergent One-Loop Diagrams}

\begin{figure}[ptb]
\includegraphics[width=6cm]{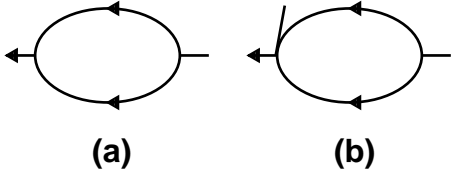}
\caption{One-loop diagrams contributing to the renormalization of
$\Gamma_{1,1}$ and $\Gamma_{1,2}$ if a momentum-cutoff regularization is used.
In dimensional regularization, these diagrams are finite.}%
\label{EinLoop}
\end{figure}

Two divergent one-loop diagrams can be assembled from the diagrammatic
elements listed in Fig.~\ref{Elemente}. These diagrams, which contribute to
the vertex functions $\Gamma_{1,1}$ and $\Gamma_{1,2}$, are shown in
Fig.~\ref{EinLoop}. It is easy to see that they are linearly divergent for
$d=d_{c}=5$ (i.e., the $\mu$ dimension is 1). No logarithmic divergencies
arise at one-loop order. Thus, the theory can be renormalized to one-loop
order by additive renormalizations
\begin{subequations}
\label{AddRen}%
\begin{align}
f  &  \rightarrow{\mathring{f}}=\mu^{\varepsilon}\bigl(v-b\Lambda
\mu^{-2\varepsilon}g\bigr)\,,\\
\tau &  \rightarrow{\mathring{\tau}}=\tau-a\Lambda\mu^{-\varepsilon}%
{\mathring{f}}=\bigl(\tau-a\Lambda v\bigr)+ab\Lambda^{2}\mu^{-2\varepsilon
}g\,,
\end{align}
where the open circles indicate unrenormalized quantities and where $a$ and $b$ are positive constants.

Next we will switch to dimensional regularization for convenience. In this
method all polynomial divergencies arising from the large cutoff are formally
set to zero, and hence the two diagrams in Fig.~\ref{EinLoop} become finite
at $d_{c}$. Thus, the additive renormalizations Eq.~(\ref{AddRen}) become
formally superfluous in dimensional regularization. However, it has to be
emphasized that dimensional regularization is only a formal trick. Physically
the additive renormalizations are always present and we need the interaction
term proportional to the coupling constant $f$ to renormalize the theory
contrary to the claim of Ref.~\cite{OhKe87}. Using dimensional regularization we have to keep in mind that these additive renormalizations do exist and that  the
critical \textquotedblleft temperature\textquotedblright\ is shifted by a term
linear in the crossover variable $v$: $\mathring{\tau}_{c}=a\Lambda
v-ab\Lambda^{2}\mu^{-2\varepsilon}g+O(g^{2})$. This $\mathring{\tau}_{c}\ $is
formally set to zero in dimensional regularization.

\subsubsection{Two-Loop calculation}

Since there are no $\varepsilon$ poles at one-loop order, we have to proceed
to higher orders in perturbation theory to find non-trivial critical
exponents. We will see that $\varepsilon$ poles do occur at two-loop order and
that the two-loop diagrams will lead us to anomalous contributions to the
critical exponents of order $\varepsilon$.

We start with the self-energy. The two-loop diagrams contributing to the
renormalization of $\Gamma_{1,1}$ are listed in Fig.~\ref{ZweiLoop1}.
\begin{figure}[ptb]
\includegraphics[width=7.5cm]{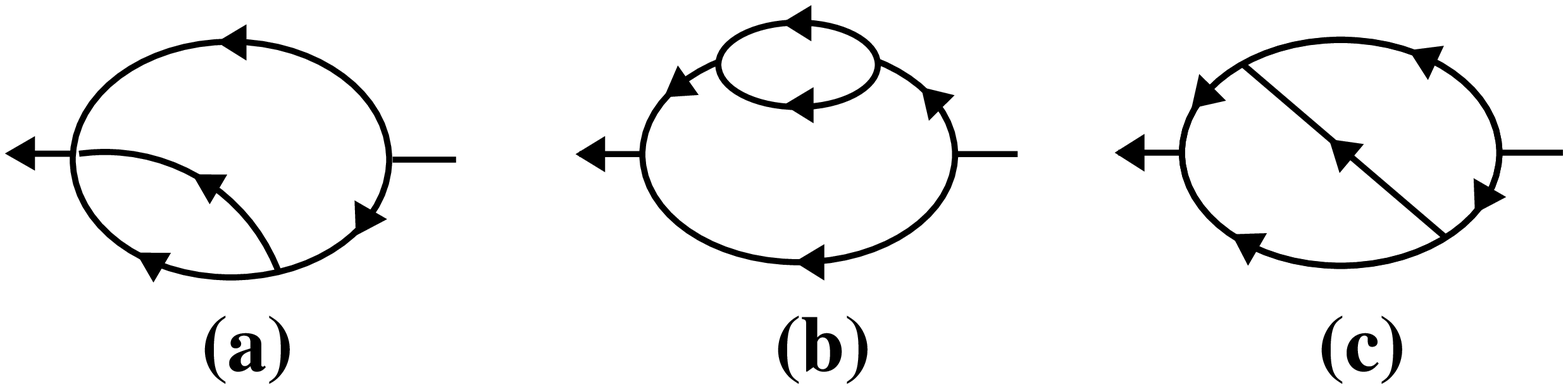}
\caption{Two-loop diagrams contributing to the renormalization of
$\Gamma_{1,1}$.}%
\label{ZweiLoop1}
\end{figure}
In the following we will use a compact notation for the diagrams. For example,
$(\ref{ZweiLoop1}a)$ refers to diagram $(a)$ of Fig.~\ref{ZweiLoop1}. For the
momentum integrals occurring in the diagrams we will use the abbreviations
\end{subequations}
\begin{equation}
I_{klm}=\int\limits_{\mathbf{q}_{1},\mathbf{q}_{2}}\frac{1}{\bigl(\mathbf{q}%
_{1}^{\,2}+1\bigr)^{k}\bigl(\mathbf{q}_{2}^{\,2}+1\bigr)^{l}\bigl((\mathbf{q}%
_{1}+\mathbf{q}_{2})^{2}+1\bigr)^{m}} \, . \label{Iklm}%
\end{equation}
This has the benefit that we can write the divergent parts of the diagrams in
a compact form. For the self-energy we have
\begin{subequations}
\label{SelbstTeile}%
\begin{align}
(\ref{ZweiLoop1}a)  &  =-\frac{g}{2}\tau^{-\varepsilon}\biggl(I_{112}%
\tau-\frac{d-4}{d}I_{113}\mathbf{q}^{2}\biggr)\, ,\\
(\ref{ZweiLoop1}b)  &  =\frac{f^{2}}{2}\tau^{-\varepsilon}I_{113}\,
,\quad(\ref{ZweiLoop1}c)=f^{2}\tau^{-\varepsilon}I_{122}\, .
\end{align}

The $I_{klm}$ can be calculated very efficient with help of the parameter
integral
\end{subequations}
\begin{equation}
I(a,b;c)=\int\limits_{\mathbf{q}_{1},\mathbf{q}_{2}}\frac{1}{\bigl(\mathbf{q}%
_{1}^{\,2}+a\bigr)\bigl(\mathbf{q}_{2}^{\,2}+b\bigr)\bigl((\mathbf{q}%
_{1}+\mathbf{q}_{2})^{2}+c\bigr)^{2}}\,. \label{parameterI}%
\end{equation}
Using dimensional regularization, we find the $\varepsilon$ expansion result
\begin{equation}
I(a,b;c)=-\frac{2\pi G_{\varepsilon}^{2}}{3\varepsilon}\,(a+b-c)+O(\varepsilon
^{0}) \label{mother}%
\end{equation}
for the parameter integral. This calculation is sketched in the Appendix. In
Eq.~(\ref{mother}) we used the shorthand $G_{\varepsilon}=\Gamma
(1+\varepsilon/2)/(4\pi)^{d/2}$ for convenience. By taking derivatives with
respect to the parameters $a$, $b$ and $c$, we get the singular parts
\begin{equation}
I_{122}=-I_{112}=-2I_{113}=\frac{2\pi G_{\varepsilon}^{2}}{3\varepsilon}
\label{IWert}%
\end{equation}
of the required original integrals. Collecting we find
\begin{align}
\Gamma_{1,1}  &  =(\tau+\mathbf{q}^{2})-(\ref{ZweiLoop1}a)-(\ref{ZweiLoop1}%
b)-(\ref{ZweiLoop1}c)\nonumber\label{Gam11}\\
&  =\biggl(1-\frac{A_{\varepsilon}g}{\varepsilon}\tau^{-\varepsilon
}\biggr)\tau-\frac{3A_{\varepsilon}f^{2}}{2\varepsilon}\tau^{-\varepsilon
}\nonumber\\
&  +\biggl(1+\frac{A_{\varepsilon}g}{10\varepsilon}\tau^{-\varepsilon
}\biggr)\mathbf{q}^{2}%
\end{align}
for the singular part of the inverse response function. Here, we introduced
$A_{\varepsilon}=\pi G_{\varepsilon}^{2}/3$ for notational convenience.

\begin{figure}[ptb]
\includegraphics[width=7.5cm]{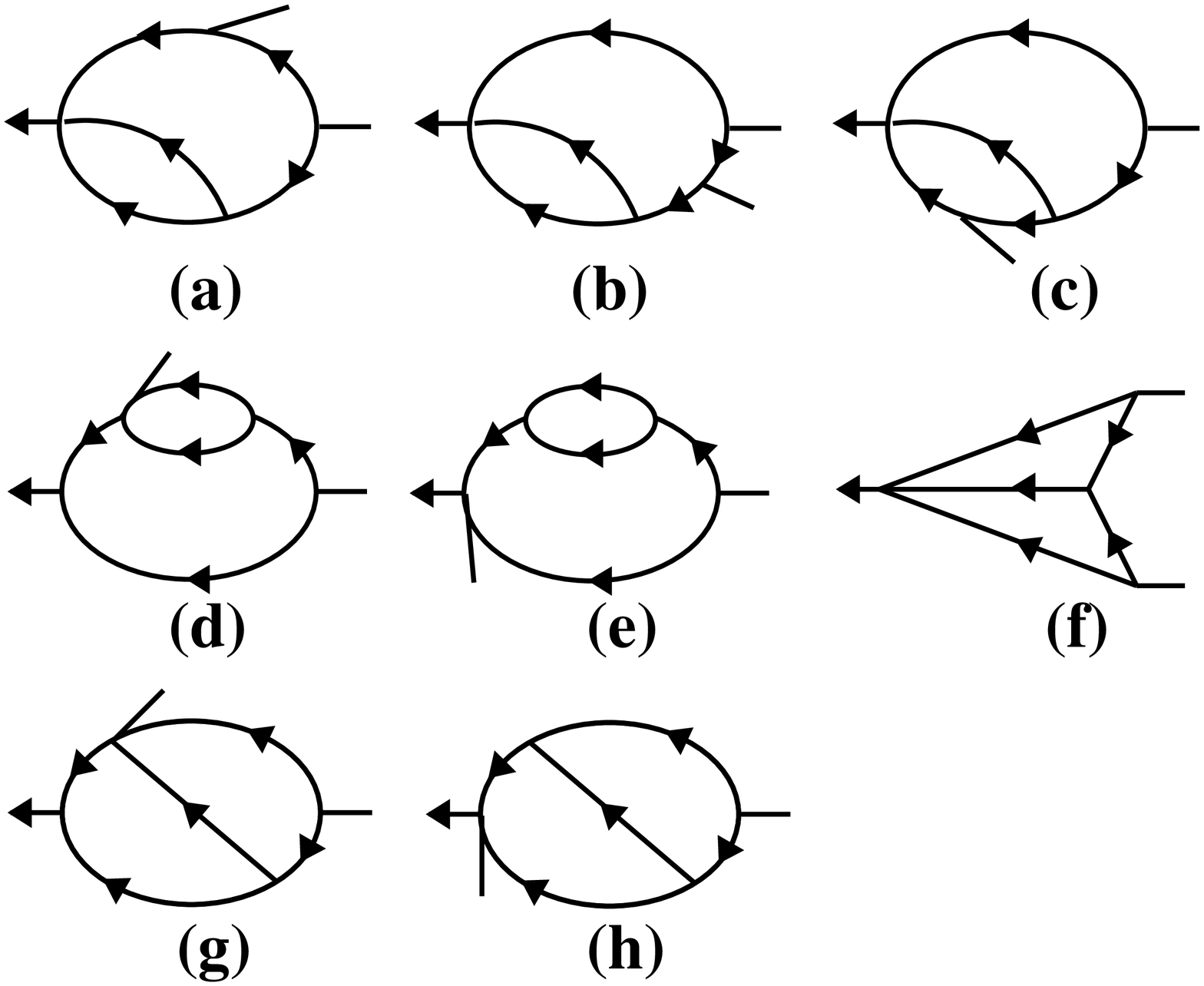}
\caption{Two-loop diagrams contributing to the renormalization of
$\Gamma_{1,2}$.}%
\label{ZweiLoop2}
\end{figure}
Now we turn to the vertex function $\Gamma_{1,2}$. Its two-loop contributions
are shown in Fig.~\ref{ZweiLoop2}. We obtain
\begin{subequations}
\label{Gam12Teile}%
\begin{align}
(\ref{ZweiLoop2}a)  &  =(\ref{ZweiLoop2}b)=(\ref{ZweiLoop2}e)=(\ref{ZweiLoop2}%
f)=fg\tau^{-\varepsilon}I_{113}\, ,\\
(\ref{ZweiLoop2}c)  &  =2\cdot(\ref{ZweiLoop2}d)=(\ref{ZweiLoop2}%
g)=(\ref{ZweiLoop2}h)=2fg\tau^{-\varepsilon}I_{122}\, .
\end{align}
Summing up the individual terms we get
\end{subequations}
\begin{equation}
\Gamma_{1,2}=\biggl(1-\frac{10A_{\varepsilon}g}{\varepsilon}\tau
^{-\varepsilon}\biggr)f\, . \label{Gam12}%
\end{equation}

\begin{figure}[ptb]
\includegraphics[width=7.5cm]{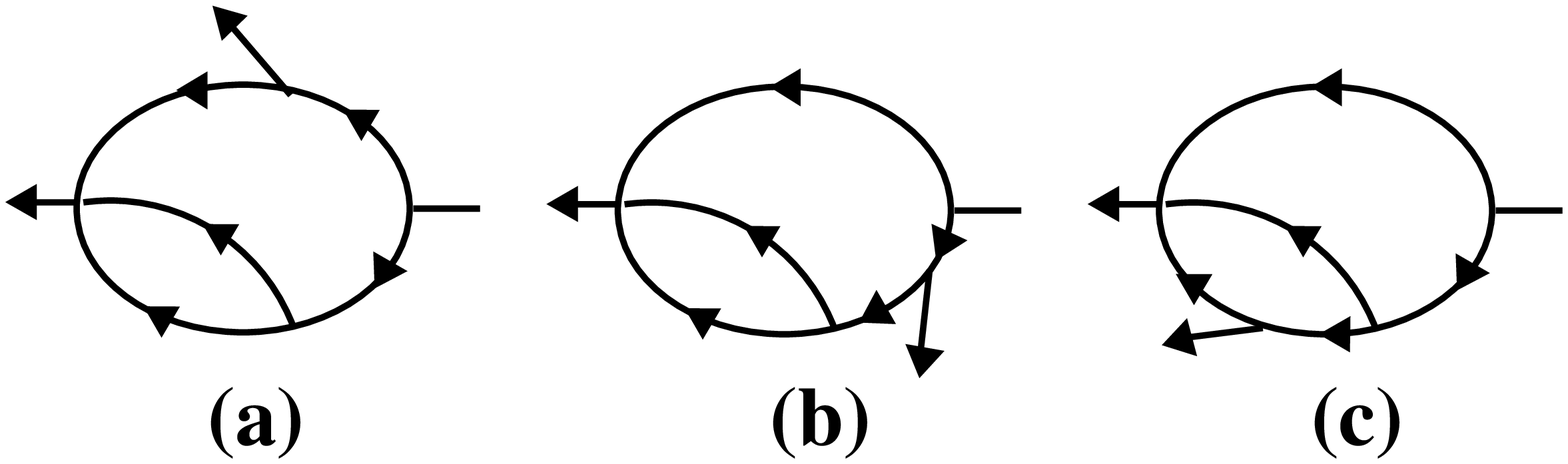}
\caption{Two-loop diagrams contributing to the renormalization of
$\Gamma_{2,1}$.}%
\label{ZweiLoop3}
\end{figure}

For the vertex function $\Gamma_{2,1}$ we have to calculate the diagrams in
Fig.~\ref{ZweiLoop3}.
With
\begin{equation}
(\ref{ZweiLoop3}a)=(\ref{ZweiLoop3}b)=-g\tau^{-\varepsilon}I_{113}\,
,\quad(\ref{ZweiLoop3}c)=-2g\tau^{-\varepsilon}I_{122}\, , \label{Gam21Teile}%
\end{equation}
we obtain
\begin{equation}
\Gamma_{2,1}=\biggl(1-\frac{2A_{\varepsilon}g}{\varepsilon}\tau^{-\varepsilon
}\biggr) \label{Gam21}%
\end{equation}

\begin{figure}[ptb]
\includegraphics[width=7.5cm]{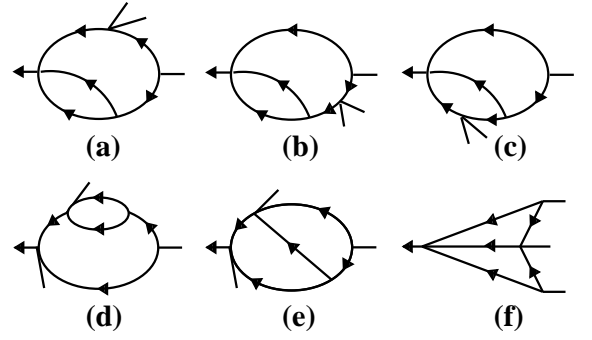}
\caption{Two-loop diagrams contributing to the renormalization of
$\Gamma_{1,3}$.}%
\label{ZweiLoop4}%
\end{figure}

It remains to consider $\Gamma_{1,3}$. The diagrams in Fig.~\ref{ZweiLoop4}
lead to
\begin{subequations}
\label{Gam13Teile}%
\begin{align}
2\cdot(\ref{ZweiLoop4}a)  &  =2\cdot(\ref{ZweiLoop4}b)=(\ref{ZweiLoop4}%
d)=3g^{2}\tau^{-\varepsilon}I_{113}\, ,\nonumber\\
2\cdot(\ref{ZweiLoop4}c)  &  =(\ref{ZweiLoop4}e)=2\cdot(\ref{ZweiLoop4}%
f)=6g^{2}\tau^{-\varepsilon}I_{122}\, ,
\end{align}
From these expressions we get
\end{subequations}
\begin{equation}
\Gamma_{1,3}=\biggl(1-\frac{18A_{\varepsilon}g}{\varepsilon}\tau
^{-\varepsilon}\biggr)g\, . \label{Gam13}%
\end{equation}

\begin{figure}[ptb]
\includegraphics[width=5.5cm]{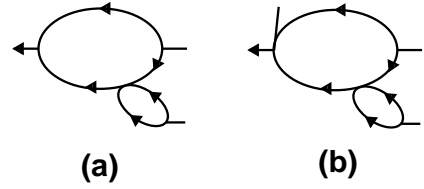}
\caption{Two-loop diagrams that contain insertions of the one-loop diagram
shown in Fig.~\ref{EinLoop}(b). In dimensional regularization, these two-loop
diagrams are finite.}%
\label{Danger}%
\end{figure}

For completeness, we conclude our quasi-static perturbation theory by briefly
returning to a cutoff regularization. In cutoff regularization there are two
additional singular two-loop diagrams, see Fig.~\ref{Danger}.
These diagrams have divergent insertions of the singular one-loop diagram
(\ref{EinLoop}b). Hence, the diagrams of Fig.~\ref{Danger} are finite in
dimensional regularization. However, in the more physical cutoff
regularization they diverge linearly with the cutoff. These divergencies are
ultimately cancelled by the additive renormalizations, Eq.~(\ref{AddRen}).

\subsection{Renormalization}

Next we absorb the $\varepsilon$-poles, into a reparametrization of the fields
and the parameters of the theory. For the quasi-static fields we employ the
renormalizations
\begin{equation}
s\rightarrow{\mathring{s}}=Z^{1/2}s\,,\quad\widetilde{s}\rightarrow
{\mathring{\widetilde{s}}}=\widetilde{Z}^{1/2}\widetilde{s}\,. \label{RenFeld}%
\end{equation}
For the parameters of the quasi-static Hamiltonian~(\ref{Hamilt}) we use the
scheme
\begin{subequations}
\label{RenPar}%
\begin{align}
&  A_{\varepsilon}g\rightarrow A_{\varepsilon}{\mathring{g}}=Z_{\lambda}%
^{-3}\widetilde{Z}Z_{u}u\mu^{2\varepsilon}\,,\\
&  A_{\varepsilon}^{1/2}f\rightarrow A_{\varepsilon}^{1/2}{\mathring{f}%
}=Z_{\lambda}^{-2}\widetilde{Z}^{1/2}Z_{v}v\mu^{\varepsilon}\,,\\
&  \tau\rightarrow{\mathring{\tau}}=Z_{\lambda}^{-1}(Z_{\tau}\tau+Yv^{2})\,,
\end{align}
where
\end{subequations}
\begin{equation}
Z_{\lambda}=(Z\widetilde{Z})^{1/2}\,. \label{Zlambd}%
\end{equation}

It follows from Eq.~(\ref{RenFeld}) that the vertex functions are renormalized
by
\begin{equation}
\Gamma_{\widetilde{N},N}\rightarrow{\mathring{\Gamma}}_{\widetilde{N}%
,N}=\widetilde{Z}^{-\widetilde{N}/2}Z^{-N/2}\,\Gamma_{\widetilde{N},N}\,.
\label{RenGam}%
\end{equation}
Using Eq.~(\ref{RenGam}) together with our two-loop results (\ref{Gam11}),
(\ref{Gam12}), (\ref{Gam21}), and (\ref{Gam13}) we find
\begin{subequations}
\label{ZFakt}%
\begin{align}
Z  &  =1-\frac{22u}{5\varepsilon}+O(u^{2})\,,\quad\widetilde{Z}=1+\frac
{21u}{5\varepsilon}+O(u^{2})\,,\label{feldFaktoren}\\
Z_{u}  &  =1+\frac{18u}{\varepsilon}+O(u^{2})\,,\quad Z_{v}=1+\frac
{10u}{\varepsilon}+O(u^{2})\,,\\
Z_{\tau}  &  =1+\frac{u}{\varepsilon}+O(u^{2})\,,\quad Y=\frac{3}%
{2\varepsilon}+O(u)\,,
\end{align}
for the renormalization factors. Equation~(\ref{feldFaktoren}) implies
\end{subequations}
\begin{equation}
Z_{\lambda}=1-\frac{u}{10\varepsilon}+O(u^{2})\,.
\end{equation}

\subsection{Renormalization group equation}

In order to explore the scaling properties of tricritical percolation we now
set up a renormalization group equation (RGE). This can be done in a routine
fashion by exploiting that the bare (unrenormalized) quantities must not
depend on the arbitrary mesoscopic length scale $\mu^{-1}$ introduced in the
course of the renormalization. In particular the bare Green functions must be
independent of $\mu$, i.e.,
\begin{equation}
\mu\left.  \partial_{\mu}\right\vert _{0}\mathring{G}_{N,\widetilde{N}}=0\,,
\label{independenceId}%
\end{equation}
where $\left.  \partial_{\mu}\right\vert _{0}$ denotes $\mu$-derivatives at
fixed bare parameters. Switching from bare to renormalized quantities, the
identity~(\ref{independenceId}) translates into the RGE
\begin{equation}
\biggl[\mathcal{D}_{\mu}+\frac{1}{2}\bigl(N\gamma+\widetilde{N}\widetilde
{\gamma}\bigr)\biggr]G_{N,\widetilde{N}}=0\,. \label{RGG}%
\end{equation}
Here, $\mathcal{D}_{\mu}$ stands for the RG differential operator
\begin{equation}
\mathcal{D}_{\mu}=\mu\partial_{\mu}+\bigl(\tau\kappa_{\tau}+v^{2}\kappa
_{v\tau}\bigr)\partial_{\tau}+v\kappa_{v}\partial_{v}+\beta_{u}\partial_{u}
\label{RGOp}%
\end{equation}
that features the Gell-Mann--Low functions
\begin{subequations}
\label{GLFu}%
\begin{align}
&  \beta_{u}=\mu\left.  \partial_{\mu}\right\vert _{0}u=\biggl(-2\varepsilon
+\frac{3}{2}\gamma+\frac{1}{2}\widetilde{\gamma}-\gamma_{u}%
\biggr)u\,,\label{betaFu}\\
&  v\kappa_{v}=\mu\left.  \partial_{\mu}\right\vert _{0}v=\biggl(-\varepsilon
+\gamma+\frac{1}{2}\widetilde{\gamma}-\gamma_{v}\biggr)v\,,\\
&  \tau\kappa_{\tau}+v^{2}\kappa_{v\tau}=\mu\left.  \partial_{\mu}\right\vert
_{0}\tau=\biggl(\frac{\gamma+\widetilde{\gamma}}{2}-\gamma_{\tau}%
\biggr)\tau-\gamma_{\tau v}v^{2}\,, \label{tauEq}%
\end{align}
and the Wilson functions
\end{subequations}
\begin{subequations}
\label{WFu}%
\begin{align}
\gamma &  =\mu\left.  \partial_{\mu}\right\vert _{0}\ln Z\,,\qquad
\widetilde{\gamma}=\mu\left.  \partial_{\mu}\right\vert _{0}\ln\widetilde
{Z}\,,\\
\gamma_{i}  &  =\mu\left.  \partial_{\mu}\right\vert _{0}\ln Z_{i}\,,\quad
i=u,v,\tau\,.
\end{align}
From Eq.~(\ref{betaFu}) we know that the functions $\beta_{u}$ and
$v\kappa_{v}$ begin with the zero-loop terms $-2\varepsilon u$ and
$-\varepsilon v$ respectively. The higher order terms are determined by the
Wilson functions. The particular form of these functions can be
straightforwardly extracted by using $\gamma_{\cdots}=\mu\left.  \partial
_{\mu}\right\vert _{0}\ln Z_{\cdots}=\beta\partial_{u}\ln Z_{\cdots}$. In
minimal renormalization the $Z$-factors have a pure Laurent expansion with
respect to $\varepsilon$, i.e., they are of the form $Z_{\cdots}=1+Z_{\cdots
}^{(1)}/\varepsilon+Z_{\cdots}^{(2)}/\varepsilon^{2}+\cdots$. Thus,
recursively in the loop expansion, the Wilson functions also have a pure
Laurent expansion. Moreover, because the Wilson functions must be finite for
$\varepsilon\rightarrow0$, all the $\varepsilon$-poles in this expansion have
to cancel (this provides a valuable check for higher
order calculations). Hence, we can obtain the Wilson functions readily from
the formula $\gamma_{\cdots}=-2u\partial_{u}Z^{(1)}$. The same argumentation
also leads to $\gamma_{\tau v}=-2Y^{(1)}-2u\partial_{u}Y^{(1)}$, where
$Y^{(1)}$ is the coefficient of the first order term in the Laurent expansion
$Y=Y^{(1)}/\varepsilon+\cdots$ of the $Y$-factor. Using this prescription, we
derive from our renormalization factors~(\ref{ZFakt}) that the Wilson
functions are given by
\end{subequations}
\begin{subequations}
\label{1LoopW}%
\begin{align}
\gamma=\frac{44u}{5}+O(u^{2})\,,  &  \qquad\widetilde{\gamma}=-\frac{42u}%
{5}+O(u^{2})\,,\\
\gamma_{u}=-36u+O(u^{2})\,,  &  \qquad\gamma_{v}=-20u+O(u^{2})\,,\\
\gamma_{\tau}=-2u+O(u^{2})\,,  &  \qquad\gamma_{\tau v}=-3+O(u)\,.
\end{align}
From these results we get
\end{subequations}
\begin{subequations}
\label{1LoopGL}%
\begin{align}
\beta_{u}  &  =\bigl(-2\varepsilon+45u+O(u^{2})\bigr)u\,,\label{betaFinal}\\
\kappa_{v}  &  =-\varepsilon+\frac{123u}{5}+O(u^{2})\,,\\
\kappa_{\tau}  &  =\frac{11u}{5}+O(u^{2})\ ,\quad\kappa_{v\tau}=3+O(u)
\end{align}
for the Gell-Mann--Low functions~(\ref{GLFu}).

\subsection{Scaling properties}
\label{staticBelow5}

\subsubsection{General scaling form}

Next we solve the RGE~(\ref{RGG}) by using the method of characteristics. The
strategy behind this method is to introduce a single flow parameter $\ell$
that allows to re-express the partial differential equation~(\ref{RGG}) as an
ordinary differential equation in terms of $\ell$. This equation then
describes how the Green functions behave under a rescaling
\end{subequations}
\begin{equation}
\mu\rightarrow\bar{\mu}(\ell)=\mu\,\ell
\end{equation}
of the inverse length scale $\mu$. The characteristic for the dimensionless
coupling constant $u$ is given by
\begin{align}
\label{uChar}\ell\frac{d \bar{u} (\ell)}{d \ell} = \beta_u( \bar{u} (\ell) ) \,
, \quad\bar{u} (1) = u \, .
\end{align}
With the exception of the characteristic for $\tau$, the remaining
characteristics are all of the same structure, viz.\
\begin{equation}
\label{Qstruct}\ell\frac{d\ln Q(\bar{u} (\ell))}{d\ell}=q(\bar{u} (\ell))\,.
\end{equation}
Here, $Q$ is a placeholder for $X$, $\widetilde{X}$, and $\bar{v}$, respectively.
$q$ is a placeholder for respectively $\gamma$, $\widetilde{\gamma}$, and
$\kappa_{v}$. The initial conditions pertaining to Eq.~(\ref{Qstruct}) are $X
(1) = \widetilde{X} (1) =1$ and $\bar{v}(1) = v$. The long-length scale behavior of
TIP corresponds to the limit $\ell\to0$. In this limit the RG flows to a
fixed point determined by the stable value $u_{\ast}$ of $u$ satisfying
$\beta_u(u_{\ast})=0$. We find that this value is given by $u_{\ast
}=2\varepsilon/45+O(\varepsilon^{2})$.

For the remainder of Sec.~\ref{staticBelow5} we exclusively consider
dimensions less than 5. We will turn to the case $d=5$ in Sec.~\ref{staticIn5}%
. In the vicinity of the fixed point $u_{\ast}$ the solution of the RGE with
the characteristics is fairly straightforward. We are confronted, however, with the
slight complication that $\tau$ itself is not a scaling variable as can be
seen from Eq.~(\ref{tauEq}). In order to diagonalize the flow equations in
near $u_{\ast}$ we switch from $\tau$ to
\begin{equation}
\sigma=\tau+\frac{\kappa_{v\tau\ast}}{\kappa_{\tau\ast}-2\kappa_{v\ast}} \,
v^{2}\, , \label{Sigma}%
\end{equation}
where $\kappa_{\tau\ast} = \kappa_{\tau}(u_{\ast})$, $\kappa_{v\ast} =
\kappa_{v} (u_{\ast})$ and so on. It can easily be checked that $\sigma$ is
governed by the flow equation
\begin{equation}
\ell\frac{d\ln\sigma(\ell)}{d\ell}= \kappa_{\tau\ast} \, ,
\end{equation}
i.e, that $\sigma$ is a true scaling variable. In the dimensions of interest
here the solutions of the characteristics for the scaling variables are of
power law form and we obtain
\begin{align}
&  G_{N,\widetilde{N}}(\{\mathbf{r}\},\tau,v,u,\mu)=\ell^{(N\eta+\widetilde{N}
\widetilde{\eta})/2}\nonumber\\
&  \qquad\qquad\times\bar{G}_{N,\widetilde{N}}(\{\mathbf{r}\},\sigma
\ell^{\kappa_{1}},v\ell^{\kappa_{2}},u_{\ast},\mu\ell)\,, \, \label{RGSol}%
\end{align}
where we have omitted nonuniversal amplitude factors. The various exponents
appearing in Eq.~(\ref{RGSol}) are given by
\begin{subequations}
\label{Exp1}%
\begin{align}
\eta &  =\gamma(u_{\ast})=\frac{88}{225}\varepsilon+O(\varepsilon^{2})\,,\\
\widetilde{\eta}  &  =\widetilde{\gamma}(u_{\ast})=-\frac{28}{75}%
\varepsilon+O(\varepsilon^{2})\,,\\
\kappa_{1}  &  =\kappa_{\tau}(u_{\ast})=\frac{22}{225}\varepsilon
+O(\varepsilon^{2})\,,\\
\kappa_{2}  &  =\kappa_{v}(u_{\ast})=\frac{7}{75}\varepsilon+O(\varepsilon
^{2})\,.
\end{align}
Supplementing the solution~(\ref{RGSol}) with a dimensional analysis to
account for naive dimensions we obtain the scaling form
\end{subequations}
\begin{align}
&  G_{N,\widetilde{N}}(\{\mathbf{r}\},\tau,v,u,\mu)=\ell^{\delta
_{N,\widetilde{N}}}\mu^{(d-4)N+2\widetilde{N}}\nonumber\label{SkalFu}\\
&  \qquad\times F_{N,\widetilde{N}}(\{\ell\mu\mathbf{r}\},\mu^{-2}\sigma
/\ell^{1/\nu},\mu^{-1}v/\ell^{\phi/\nu})\,,
\end{align}
where
\[
\delta_{N,\widetilde{N}}=\Bigl(d-4+\frac{\eta}{2}\Bigr)N+\Bigl(2+\frac
{\widetilde{\eta}}{2}\Bigr)\widetilde{N}\,,
\]
and where the $F_{N,\widetilde{N}}$ are, up to nonuniversal amplitude factors,
universal scaling functions. The scaling exponents of the correlation length
and the crossover variable are
\begin{subequations}
\label{Exp2}%
\begin{align}
\nu &  =\frac{1}{2-\kappa_{1}}=\frac{1}{2}+\frac{11}{450}\varepsilon
+O(\varepsilon^{2})\,,\\
\phi &  =\frac{1-\kappa_{2}}{2-\kappa_{1}}=\frac{1}{2}-\frac{1}{45}%
\varepsilon+O(\varepsilon^{2})\,.
\end{align}

In the following we will concentrate on a path in the phase diagram spanned by
the relevant variables $\tau$ and $v$ which approaches the tricritical point
$\tau=v=0$ as $v\sim\tau$. Hence, we will neglect the crossover variable
$v/\left\vert \sigma\right\vert ^{\phi}\sim\left\vert \tau\right\vert
^{1-\phi}\ll1$, where we have set $\mu=1$ for convenience. In this regime we
can write the fundamental scaling form~(\ref{SkalFu}) as
\end{subequations}
\begin{equation}
G_{N,\widetilde{N}}(\{\mathbf{r}\},\tau)=\left\vert \tau\right\vert
^{N\beta+\widetilde{N}{\beta}^{\prime}}F_{N,\widetilde{N}}^{\pm}(\{\left\vert
\tau\right\vert ^{\nu}\mathbf{r}\})\,, \label{SkalGreen}%
\end{equation}
with
\begin{subequations}
\label{beta}%
\begin{align}
\beta &  =\nu\Bigl(d-4+\frac{\eta}{2}\Bigr)=\frac{1}{2}-\frac{17}%
{45}\varepsilon+O(\varepsilon^{2})\,,\\
{\beta}^{\prime}  &  =\nu\Bigl(2+\frac{\widetilde{\eta}}{2}\Bigr)=1-\frac
{2}{45}\varepsilon+O(\varepsilon^{2})\,.
\end{align}
The superscript of the scaling functions $F_{N,\widetilde{N}}^{\pm}$
corresponds to the sign of $\tau$. Note that $\xi\sim\left\vert \tau
\right\vert ^{-\nu}$ is the correlation length.

\subsubsection{Scaling behavior of various percolation observables}

Now we will exploit our knowledge about the correlation functions of the
fields to extract the scaling behavior of various observables that play an
important role in percolation theory. First we will consider the case that the
process starts with a single seed at the origin $\mathbf{r}=0$. Second we will
look at the case that the density of the initial state is homogeneous.

Let us start by considering clusters of a given finite size $S$,
i.e.\ clusters with a finite mass $S$ of the debris given that the
process started with a seed, a single agent, at the origin $\mathbf{r}=0$. In principle, any reasonable initial state can be prepared by choosing the appropriate seed density $\rho_{0}(\mathbf{r})$. In the Langevin equation~(\ref{eq:General Langevin}) this
general initial condition corresponds to an additional source term $\lambda^{-1}%
{q}(\mathbf{r},t)=\rho_{0}(\mathbf{r})\,\delta(\lambda t)$. At the level of
the dynamic response functional $\mathcal{J}$ (\ref{J}) such an initial state
translates into a further additive contribution $-\int
d^{d}x\,dt\,q(\mathbf{r},t)\tilde{n}(\mathbf{r},t)$. Thus, a seed
$q(\mathbf{r},t)=\delta(\mathbf{r})\delta(t)$ is represented by the
contribution $-\tilde{n}(\mathbf{0},0)$. At the level of the quasistatic
Hamiltonian $\mathcal{H}$ (\ref{Hamilt}) such a seed is therefore represented by an
additive term $-\tilde{s}(0)$.

Let $P(S)dS$ be the measure for the probability that the cluster mass of the
debris generated by a seed at the origin is between $S$ and $S+dS$. In our field theoretic formulation the 
probability density $P(S)$ can be expressed as
\end{subequations}
\begin{equation}
P(S)=\left\langle \delta\Bigl(\int d^{d}r\,s(\mathbf{r})-S\Bigr)\,\exp
\left[ \tilde{s}(\mathbf{0}) \right] \right\rangle \, . \label{P}%
\end{equation}
For big clusters with $S\gg1$ we can expand the exponential to first order (higher orders lead
asymptotically only to subleading corrections) and obtain%
\begin{equation}
P_{\text{as}}(S)=\left\langle \delta\Bigl(\int d^{d}r\,s(\mathbf{r}%
)-S\Bigr)\,\,\tilde{s}(\mathbf{0})\right\rangle \,  \label{P-as}%
\end{equation}
for the asymptotic distribution. We will return to $P_{\text{as}}(S)$ in a moment.

The percolation probability $\mathcal{P}_{\infty}$ is defined as the
probability for the existence of an infinite cluster generated from a single seed.
Hence $\mathcal{P}_{\infty}$ is given by
\begin{align}
\mathcal{P}_{\infty}  &  =1-\lim_{c\rightarrow+0}\int_{0}^{\infty
}dS\,\mathrm{e}^{-cS}P(S)\nonumber\\
&  =1-\lim_{c\rightarrow+0}\langle\exp\Big(\widetilde{s}(\mathbf{0})-c\int
d^{d}r\,s(\mathbf{r})\Big)\rangle. \label{Probinf1}%
\end{align}
Via expanding $\exp [ \tilde{s}(\mathbf{0}) ]$ we obtain the asymptotic
form \cite{Ja03}
\begin{equation}
\mathcal{P}_{\infty}\simeq-\lim_{c\rightarrow+0}\langle\tilde{s}%
(\mathbf{0})\mathrm{e}^{-cM}\rangle\,, \label{PerkProb}%
\end{equation}
where $M=\int d^{d}x\,s(\mathbf{x})$. The virtue of this formula is that it
relates the percolation probability unambiguously to an expression accessible
by field theory. For actual calculations the term $\exp(-cM)$ has to be
incorporated into the quasi-static Hamiltonian. This leads to
\begin{equation}
\mathcal{H}_{c}=\mathcal{H}+\int d^{d}r\,c(\mathbf{r})s(\mathbf{r})
\end{equation}
instead of the original $\mathcal{H}$. Here, $c(\mathbf{x})=c$ is a source
conjugate to the field $s$. Whereas in general $\langle\tilde{s}\rangle=0$ if $c=0$ by virtue of
causality, the limit $c\rightarrow+0$ leads to a nonvanishing order
parameter $\mathcal{P}_{\infty}$ in the spontaneously symmetry broken active phase. Having introduced $\mathcal{H}_{c}$,
we can write
\begin{align}
\mathcal{P}_{\infty}  &  =-\lim_{c\rightarrow+0}\langle\tilde{s}%
(\mathbf{0})\rangle_{c}\nonumber\\
&  =-G_{0,1}(\mathbf{0},\tau,c\rightarrow+0,u,\mu)\,, \label{PerkProb1}%
\end{align}
where $\langle\cdots\rangle_{c}$ denotes averaging with respect to
$\mathcal{H}_{c}$. With the help of Eq.~(\ref{PerkProb1}) and the scaling
form~(\ref{SkalGreen}) we readily obtain that
\begin{equation}
\mathcal{P}_{\infty}\sim\theta(-\tau)\left\vert \tau\right\vert ^{\beta
^{\prime}}\,. \label{Probinf}%
\end{equation}

In order to examine the scaling behavior of $P(S)$ we can look at its moments
defined by
\begin{equation}
\langle S^{k}\rangle=\int_{0}^{\infty}dS\,S^{k}P(S)\,.
\end{equation}
Using Eq.~(\ref{P-as}), our scaling result~(\ref{SkalGreen}) leads to
\begin{equation}
\langle S^{k}\rangle\simeq\int(d^{d}r)^{k}\,G_{k,1}(\{\mathbf{r}%
\},\mathbf{0},\tau)\sim\left\vert \tau\right\vert ^{{\beta}^{\prime}%
-k(d\nu-\beta)}\,. \label{Moments}%
\end{equation}
This tells us that $P_{\text{as}}$ scales as
\begin{equation}
P_{\text{as}}(S,\tau)=S\,n_{S}(\tau)=S^{1-\tau_{p}}f(\tau S^{\sigma_{p}})\,,
\label{Pskal}%
\end{equation}
where $n_{S}$ is the number of clusters of size $S$ per lattice site. The
$n_{S}$ play an important role in percolation theory where they are called
cluster numbers. The scaling exponents in the scaling form~(\ref{Moments})
reflect the usual nomenclature of percolation theory. They are given by
\begin{subequations}
\label{ClNumbScal}%
\begin{align}
\sigma_{p}  &  =\frac{1}{d\nu-\beta}=\frac{1}{2}+O(\varepsilon^{2})\,,\\
\tau_{p}  &  =\frac{{\beta}^{\prime}}{d\nu-\beta}+2=\frac{5}{2}-\frac{1}%
{45}\varepsilon+O(\varepsilon^{2})\,.
\end{align}
These exponents coincide with the corresponding exponents of conventional
isotropic percolation only in mean-field theory. It follows from
Eq.~(\ref{Moments}) that the mean cluster mass $\langle S\rangle$ of the
finite clusters scales as
\end{subequations}
\begin{equation}
\langle S\rangle=\mathcal{M}(\tau)=\mathcal{M}_{0}\,\left\vert \tau\right\vert
^{-\gamma}, \label{MeanMass}%
\end{equation}
with the exponent
\begin{equation}
\gamma=d\nu-(\beta+{\beta}^{\prime})=1+\frac{2}{45}\varepsilon+O(\varepsilon
^{2})\,. \label{gammaExponent}%
\end{equation}

Next we consider Green functions restricted to clusters of given mass
$S$. In terms of the conventional unrestricted averages with respect to
$\mathcal{H}$, these restricted Green functions can be expressed for
large $S$ as
\begin{align}
&  C_{N}^{(S)}(\{\mathbf{r}\},\tau)\nonumber\\
&  =\left\langle s(\mathbf{r}_{1})\cdots s(\mathbf{r}_{N})\delta\Bigl(\int
d^{d}r\,s(\mathbf{r})-S\Bigr)\widetilde{s}(\mathbf{0})\right\rangle
^{(\text{conn})}\,.
\end{align}
Equation (\ref{SkalGreen}) leads to the scaling form
\begin{equation}
C_{N}^{(S)}(\{\mathbf{r}\},\tau)=\left\vert \tau\right\vert ^{(N-1)\beta
+\beta^{\prime}+d\nu}F_{N}(\{\left\vert \tau\right\vert ^{\nu}\mathbf{r}%
\},\left\vert \tau\right\vert ^{d\nu-\beta}S)\,. \label{KorrelFu}%
\end{equation}
With help of these restricted Green functions we can write the radius of
gyration (mean-square cluster radius) of clusters of size $S$ as
\begin{equation}
R_{S}^{2}=\frac{\int d^{d}r\,\mathbf{r}^{2}C_{1}^{(S)}(\mathbf{r,\tau}%
)}{2d\int d^{d}r\,C_{1}^{(S)}(\mathbf{r,\tau})}\,. \label{R_s}%
\end{equation}
Equation~(\ref{KorrelFu}) then leads to
\begin{equation}
R_{S}^{2}=S^{2/D_{f}}f_{R}(\tau S^{\sigma_{p}})\ \label{GyrRad}%
\end{equation}
with the fractal dimension
\begin{equation}
D_{f}=d-\frac{\beta}{\nu}=4-\frac{44}{225}\varepsilon+O(\varepsilon^{2})\ .
\label{FraktDim}%
\end{equation}

We conclude Sec.~\ref{staticGGEP} by considering the scaling behavior of the
statistics of the debris if the initial state is prepared with a
homogeneous seed density $\rho_{0}$. As discussed above, such an initial state
translates at the level of the quasistatic Hamiltonian $\mathcal{H}$~(\ref{Hamilt}) into a further additive contribution $-\rho_{0}\int d^{d} r\,\widetilde{s}(\mathbf{r})$. Our general scaling form~(\ref{SkalGreen})
implies that the correlation functions of the densities $s(\mathbf{r})$ scale
in case of the homogeneous initial condition like
\begin{align}
G_{N,\tilde{N}}(\{\mathbf{r}\},\tau,\rho_{0})  &  =\sum_{k=0}^{\infty}%
\frac{\rho_{0}^{k}}{k!}\int(d^{d}\widetilde{r})^{k}\,G_{N,\widetilde{N}%
+k}(\{\mathbf{r}\},\{\widetilde{\mathbf{r}}\},\tau)\nonumber\\
&  =\left\vert \tau\right\vert ^{\beta N+\beta^{\prime}\tilde{N}}F_{N}^{\pm
}(\{\left\vert \tau\right\vert ^{\nu}\mathbf{r}\},\left\vert \tau\right\vert
^{{\beta}^{\prime}-d\nu}\rho_{0})\ . \label{KorFuhom2}%
\end{align}
It is obvious that the initial seed density plays the role of an ordering
field. Hence, the Green functions do not show critical singularities as long
as $\rho_{0}$ is finite. For the homogeneous initial condition the appropriate
order parameter is given by the density of the debris
\begin{equation}
\rho=\langle s(\mathbf{r})\rangle_{\rho_{0}}=G_{1,0}(\mathbf{r},\tau,\rho
_{0})\,.
\end{equation}
Equation~(\ref{KorFuhom2}) tells us that
\begin{equation}
\rho=\left\vert \tau\right\vert ^{\beta}f_{\rho}^{\pm}(\left\vert
\tau\right\vert ^{{\beta}^{\prime}-d\nu}\rho_{0})\,. \label{Dichte}%
\end{equation}
In the non-percolating phase ($\tau>0$) the order parameter $\rho$ is linear
in $\rho_{0}$ for small seed density with a susceptibility coefficient that
diverges as $\tau\rightarrow0$,
\begin{equation}
\rho(\tau>0,\rho_{0})\sim\tau^{-\gamma}\rho_{0}\,, \label{nDichte}%
\end{equation}
with the susceptibility exponent $\gamma$ given in Eq.~(\ref{gammaExponent}).
At criticality ($\tau=0$) the order parameter $\rho$ goes to zero for
$\rho_{0}\rightarrow0$ as
\begin{equation}
\rho(\tau=0,\rho_{0})\sim\rho_{0}^{1/\delta}\,, \label{cDichte}%
\end{equation}
with the exponent
\begin{equation}
\delta=\frac{d\nu-{\beta}^{\prime}}{\beta}=1+\frac{\gamma}{\beta}=3+\frac
{8}{5}\varepsilon+O(\varepsilon^{2})\,.
\end{equation}
Finally, in the percolating phase ($\tau<0$) the order parameter becomes
independent of the initial seed density in the limit $\rho_{0}\rightarrow0$
and goes to zero with $\tau$ as
\begin{equation}
\rho(\tau<0)\sim\left\vert \tau\right\vert ^{\beta}\,. \label{pDichte}%
\end{equation}

Equations~(\ref{Probinf}) and (\ref{pDichte}) show explicitly that the two
order parameters, namely the density $\rho$ of the debris and the percolation
probability $\mathcal{P}_{\infty}$, have different exponents $\beta$ and
${\beta}^{\prime}$. This stands in contrast to ordinary isotropic as well as
directed percolation. As mentioned earlier, ordinary isotropic and directed percolation are special
in the sense that they posses an asymptotic symmetry that leads to the
equality of the respective exponents $\beta$ and ${\beta}^{\prime}$.

\subsection{Scaling properties in 5 dimensions -- logarithmic corrections}

\label{staticIn5}

Here we will study TIP directly in 5 dimensions where fluctuations induce logarithmic corrections to
the leading mean-field terms rather than anomalous exponents. First, we will
establish the general form of the logarithmic corrections. Second, we will
analyze how these corrections, to leading order, affect the observables
studied in Sec.~\ref{staticBelow5}

\subsubsection{General form of the logarithmic corrections}

\label{genFormStaic} Now we will solve the characteristics directly for $d=5$.
The Wilson functions stated in Eqs.~(\ref{1LoopW}) and (\ref{1LoopGL}) are
central ingredients of the characteristics. For ecconomic reasons, we will use
in the following an abbreviated notion for the Wilson functions of the type
$f(u) = f_{1} u + f_{2} u^{2} + \cdots$. For instance, we will write
Eq.~(\ref{betaFinal}) as $\beta_u(u) = \beta_{1} u + \beta_{2} u^{2} + O(u^{3})$
and likewise for the other Wilson functions. Since we are interested in the
tricritical point, we set $v=0$ throughout.

Solving the characteristic for $u$, Eq.~(\ref{uChar}), for $\varepsilon=0$
yields readily
\begin{equation}
\ell=\ell(w)=\ell_{0}\exp\bigg[-\frac{1}{\beta_{2}w}+O(\ln w)\bigg]\,,
\label{l(w)}%
\end{equation}
where $\ell_{0}$ is an integration constant and where we abbreviated
\begin{equation}
w=\bar{u}\,.
\end{equation}
To leading order we obtain from Eq.~(\ref{l(w)}) asymptotically for $\ell\ll1$%
\begin{equation}
w\simeq\frac{1}{45\left\vert \ln\ell\right\vert }\ll1. \label{w-as}%
\end{equation}
The remaining characteristics, Eq.~(\ref{Qstruct}) are solved via exploiting
$\ell d/d\ell=\beta d/dw$ with the asymptotic result
\begin{equation}
Q(w)\simeq Q_{0}\,w^{q_{1}/\beta_{2}}\,, \label{Q(w)}%
\end{equation}
where $Q_{0}$ symbolizes a non-universal integration constant. Having solved
the characteristics, we obtain
\begin{align}
G_{N,\widetilde{N}}(\left\{  \mathbf{r}\right\}  ,\tau)  &  =[\mu
\ell\,X(w)^{1/2}]^{N}[(\mu\ell)^{2}\tilde{X}(w)^{1/2}]^{\widetilde{N}%
}\nonumber\\
&  \times F_{N,\widetilde{N}}\Big(\left\{  \mu\ell\mathbf{r}\right\}
,\frac{\tau(w)}{(\mu\ell)^{2}},w\Big) \label{SolG-gen}%
\end{align}
as a formal scaling solution for the Green functions.

In the following we have to be careful about the explicit dependence of the scaling
functions $F_{N,\widetilde{N}}$ on $w$ even though we are only interested in
the leading logarithmic corrections. This intricacy comes from the fact that rather mean-field than Gaussian theory applies at the upper critical dimension. Hence we must carefully distinguish between the 2 roles of $g$, viz.\ its role  as a dangerous irrelevant
variable which scales the fields and the parameters in the correlation and
response functions, and, its role (in its dimensionless form $u$) as the loop expansion
parameter. Only the latter role can be safely neglected for the leading
logarithmic corrections. As we shall see in the following, the dependence of the functions
$F_{N,\widetilde{N}}$ on $w$ if given by
\begin{align}
& F_{N,\widetilde{N}}\left(  \left\{  \mathbf{r}\right\}  ,\tau,v,w\right)   
\nonumber \\
& =w^{(1-\tilde{N})/2}\Bigl[ F_{N,\widetilde{N}}^{\prime}\left(  \left\{
\mathbf{r}\right\}  ,\tau,v/\sqrt{w}\right) +O(\sqrt{w})\Bigr]. \label{Dangerous}%
\end{align}
To derive this result one can consider the loop-scaling of the generating
functionals of the vertex and Green functions as it was done in
Ref.~\cite{JaSt04} for studying logarithmic corrections in DP. Here we use a more direct method to determine the dependence of the tree diagrams on the dangerous variable $g$. Consider an
arbitrary connected diagram with $L$ loops, $P$ propagators, $N$ external
$s$-legs, $\tilde{N}$ external $\tilde{s}$-legs, $V_{g}$ vertices of type $g$,
$V_{f}$ vertices of type $f$, and $V_{1}$ noise vertices proportional to $1$.
This generic diagram contributing to the Green function $G_{N,\tilde{N}}$ satisfies the three topological conditions
\begin{subequations}
\label{Conditions}%
\begin{align}
L  &  =P-V_{g}-V_{f}-V_{1}-N-\tilde{N}+1,\\
P  &  =3V_{g}+2V_{f}+V_{1}+N,\\
P  &  =V_{g}+V_{f}+2V_{1}+\tilde{N}.
\end{align}
Eliminating $P$ we arrive at two equations for the number of vertices, namely
\end{subequations}
\begin{subequations}
\label{VertNr}%
\begin{align}
V_{1}  &  =N-1+L,\label{VertNr1}\\
2V_{g}+V_{f}  &  =\tilde{N}-1+L. \label{VertNr2}%
\end{align}
Switching from $f$ to the scaled variable $f^{\prime}=f/\sqrt{g}$, we
find that the diagram scales with $g$ as $g^{(\tilde{N}-1+L)/2}$ with the factor
$g^{L/2}$ being determined by the loop-order of the diagram. Upon renormalization this reasoning leads for $L=0$ to the scaling form~(\ref{Dangerous}).

Knowing Eq.~(\ref{Dangerous}) we can write down the scaling form of the Green functions with its leading logarithmic corrections. From the general solution of the RGE for the Green functions~(\ref{SolG-gen}) in conjunction with the asymptotic result (\ref{w-as}) we get \end{subequations}
\begin{align}
G_{N,\widetilde{N}}(\left\{  \mathbf{r}\right\}  ,\tau)  &  =\ell
^{N+2\tilde{N}}\left\vert \ln\ell\right\vert ^{7\tilde{N}/75-22N/225+(1-\tilde
{N})/2}\nonumber\\
&  \times F_{N,\tilde{N}}\bigl(\{\ell\mathbf{r}\},\ell^{-2}\left\vert \ln
\ell\right\vert ^{-11/225} \tau \bigr). \label{StatLogSkal}%
\end{align}

\subsubsection{Logarithmic corrections to the percolation observables}

As above we will first consider the case that the process emanates from a
single local seed at the origin. Exploiting Eq.~(\ref{PerkProb1}), we find
that the percolation probability has the scaling form
\begin{equation}
\mathcal{P}_{\infty}=\ell^{2}\left\vert \ln\ell\right\vert ^{7/75} f_{\mathcal{P}_{\infty}} (\ell
^{-2}\left\vert \ln\ell\right\vert ^{-11/225}\tau),
\end{equation}
where $f_{\mathcal{P}_{\infty}}$ is non zero only for $\tau<0$. Now we fix the arbitrary but small
flow parameter $\ell$ so that $\tau^{-1}$ effectively acquires a finite value
in the scaling limit,
\begin{equation}
\ell^{-2}\left\vert \ln\ell\right\vert ^{-11/225}\tau\sim1. \label{choiceTau}%
\end{equation}
Hence, we choose asymptotically%
\begin{equation}
\ell^{2}\sim\left\vert \tau\right\vert \left\vert \ln\left\vert \tau
\right\vert \right\vert ^{-11/225}. \label{lTau}%
\end{equation}
Collecting, we obtain that
\begin{equation}
\mathcal{P}_{\infty}(\tau)\sim\theta(-\tau)\left\vert \tau\right\vert
\left\vert \ln\left\vert \tau\right\vert \right\vert ^{2/45}.
\label{LogP_inft}%
\end{equation}

Next we consider the probability $P(S,\tau)$. Using the definition
(\ref{Moments}) and the general result (\ref{StatLogSkal}) we obtain
\begin{align}
P(S,\tau) &  =\ell^{6}\left\vert \ln\ell\right\vert ^{43/225}\nonumber\\
&  \times f_P^\prime \bigl(\ell^{-2}\left\vert \ln\ell\right\vert ^{-11/225}\tau
,\ell^{4}\left\vert \ln\ell\right\vert ^{22/225}S\bigr)\,.\label{P-lSkal}%
\end{align}
Being interested primarily in $P(S,\tau)$ as a function near criticality
$\tau\approx0$ and not in the animal limit $S\rightarrow\infty$ with $\tau>0$,
we hold the second argument finite. Hence we choose
\begin{equation}
\ell^{4}\sim S^{-1}(\ln S)^{-22/225}\,.\label{choiceS}%
\end{equation}
With these settings we find
\begin{equation}
P(S,\tau)=S^{-3/2}\,(\ln S)^{2/45}\,f_{P}\bigl(\tau S^{1/2}%
\bigr).\label{P-logSkal}%
\end{equation}
The scaling function $\,f_{P}\left(  x\right) $ is found with help of a simple mean
field calculation to be proportional to $\exp(-\mathrm{const} \, x^2)\,$as long
as $x$ is finite. For $x\rightarrow\infty$ it crosses over to the animal limit
with its own independent scaling behavior. Note that there is, to the order we
are working, no logarithmic correction associated with the $S^{1/2}$ in the
argument of the scaling function. As a corollary of Eq.~(\ref{P-logSkal}), we
obtain the critical behavior of the mean cluster mass,
\begin{equation}
\mathcal{M(\tau)}=\langle S\rangle\sim\tau^{-1}\,\left\vert \ln\tau\right\vert
^{2/45}.\label{meanClM}%
\end{equation}

Now we turn to clusters of a given size $S$. The scaling behavior of the Green
functions restricted to these clusters takes on the form
\begin{align}
&  C_{N}^{(S)}(\{\mathbf{r}\},\tau)=\ell^{N+6}\left\vert \ln\ell\right\vert
^{(43-22N)/225}\nonumber\label{motherR}\\
&  \times F_{N}\bigl(\left\{  \ell\mathbf{r}\right\}  ,\ell^{-2}\left\vert
\ln\ell\right\vert ^{-11/225}\tau,\ell^{4}\left\vert \ln\ell\right\vert
^{22/225}S\bigr)
\end{align}
in 5 dimensions. From this scaling it is straightforward to extract the radius
of gyration of clusters of size $S$ via definition~(\ref{R_s}). Choosing
$\ell$ per Eq.~(\ref{choiceS}) leads to
\begin{equation}
R_{S}^{2}=S^{1/2}\,(\ln S)^{11/225}\,f_{R}\bigl(\tau S^{1/2}\bigr).
\end{equation}
In mean-field theory, as a straightforward calculation shows, the scaling function $f_{R}(x)$ is identical to a non-universal constant.

In the remainder of Sec.~\ref{staticIn5} we will consider the homogeneous
initial condition. The $N$-point density correlation functions obey in $d=5$
the scaling form
\begin{align}
&  G_{N,\tilde{N}}(\{\mathbf{r}\},\tau,\rho_{0})=\ell\,^{N+2\tilde{N}%
}\left\vert \ln\ell\right\vert ^{1/2+61\tilde{N}/150-22N/225}\nonumber\\
&  \times F_{N,\tilde{N}}\bigl(\left\{  \ell\mathbf{r}\right\}  ,\ell
^{-2}\left\vert \ln\ell\right\vert ^{-11/225}\tau,\ell^{-3}\left\vert \ln
\ell\right\vert ^{-61/150}\rho_{0}\bigr).\label{KorFuhom3}%
\end{align}
Specifying Eq.~(\ref{KorFuhom3}) to $N=1$ and $\tilde{N}=0$, we obtain the
scaling behavior of the mean density of the debris,
\begin{align}
&  \bar{\rho}(\tau,\rho_{0})=\ell\left\vert \ln\ell\right\vert ^{181/450}%
\nonumber\\
&  \times F_{1,0}\bigl(\ell^{-2}\left\vert \ln\ell\right\vert ^{-11/225}%
\tau,\ell^{-3}\left\vert \ln\ell\right\vert ^{-61/150}\rho_{0}%
\bigr).
\label{rho-lSkal}%
\end{align}
For $\tau\neq0$ and small seed density $\rho_{0}$ it is appropriate to choose
$\ell$ according to Eq.~(\ref{choiceTau}). This provides us with
\begin{equation}
\bar{\rho}(\tau,\rho_{0})=\left\vert \tau\right\vert ^{1/2}\,\left\vert
\ln\left\vert \tau\right\vert \right\vert ^{17/45}\,f_{\bar{\rho}}^{\pm
}\bigl(\rho_{0}\,\left\vert \tau\right\vert ^{-3/2}\,\left\vert \ln\left\vert
\tau\right\vert \right\vert ^{-1/3}\bigr)\,.\label{meanDens5}%
\end{equation}
The scaling function $f_{\bar{\rho}}^{+}(x)$ behaves like $f_{\bar{\rho}}^{+}(x)\sim x$ for small $x$, and hence
\begin{equation}
\bar{\rho}(\tau>0,\rho_{0})\sim\rho_{0}\,\tau^{-1}\,\left\vert \ln
\tau\right\vert ^{2/45}%
\end{equation}
in this regime. At $\tau=0$, the mean density is a function of $\rho_{0}$
only. To obtain the logarithmic corrections for this case, we set
\begin{equation}
\ell\sim\rho_{0}\,^{1/3}\,\left\vert \ln\rho_{0}\right\vert ^{-61/450}%
.\label{choiceRho0}%
\end{equation}
From Eq.~(\ref{rho-lSkal}) in conjunction with Eq.~(\ref{choiceRho0}) we
obtain
\begin{equation}
\bar{\rho}(\tau=0,\rho_{0})\sim\rho_{0}^{1/3}\,\left\vert \ln\rho
_{0}\right\vert ^{4/15}\,.\label{StatDens}%
\end{equation}
Lastly, the order parameter should be independent of the seed density in the
percolating phase for $\rho_{0}\rightarrow0$, i.e., the scaling function
$f_{\bar{\rho}}^{-}$ of Eq.~(\ref{meanDens5}) should approach a constant in
this limit. Thus, we get
\begin{equation}
\bar{\rho}(\tau<0)\sim\left\vert \tau\right\vert ^{1/2}\,\left\vert
\ln\left\vert \tau\right\vert \right\vert ^{17/45}%
\end{equation}
in this regime.

\section{Dynamic Scaling Properties}

\label{dynGGEP}


In this section we will study dynamic scaling properties of TdIP. To find out these  properties we have to renormalize the dynamic response
functional~(\ref{J}). In comparison to the quasi-static
Hamiltonian~(\ref{Hamilt}), $\mathcal{J}$ has one additional parameter, namely
the kinetic coefficient $\lambda$. To determine the renormalization of
$\lambda$, we have to calculate one of the dynamic vertex functions in full
time or frequency dependence. A dynamic RGE then leads to general scaling form
for the dynamic Green functions. This scaling form allows us to deduce the
dynamic scaling behavior of various percolation observables.

\subsection{Diagrammatics}

\begin{figure}[ptb]
\includegraphics[width=5.5cm]{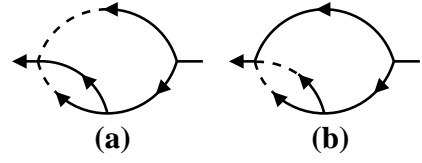}
\caption{Dynamic two-loop diagrams which provide us with the renormalization
factor of the kinetic coefficient $\lambda$.}%
\label{DynGraph}%
\end{figure}

In order to determine the renormalization of $\lambda$, we could calculate the
frequency dependent part of any of the superficially divergent vertex
functions. For convenience, we choose to work with $\Gamma_{1,1}$. Two dynamic
two-loop diagrams for the self-energy can be constructed from the diagrammatic
elements in Fig.~\ref{fullDynElements}. These two diagrams are shown in
Fig.~\ref{DynGraph}.
After Fourier transformation and some rearrangements, we get from the diagrams
the contribution
\begin{equation}
(\ref{DynGraph}a)+(\ref{DynGraph}b)=i\omega g\tau^{-\varepsilon}%
\biggl(-\frac{B}{8}+\frac{C}{2}+\frac{3}{8}I_{122}+\frac{3}{4}I_{113}%
\biggr) \label{dynSelf}%
\end{equation}
to $\Gamma_{1,1}$, where we have not displayed various terms (those not linear
in the frequency $\omega$) for notational simplicity. $B$ and $C$ stand for
the integrals
\begin{equation}
B=\int\limits_{\mathbf{q}_{1},\mathbf{q}_{2}}\frac{1}{\bigl(\mathbf{q}%
_{1}^{\,2}+1\bigr)^{2}\bigl(\mathbf{q}_{2}^{\,2}+1\bigr)^{2}\bigl(\mathbf{q}%
_{1}^{\,2}+\mathbf{q}_{2}^{\,2}+(\mathbf{q}_{1}+\mathbf{q}_{2})^{2}+3\bigr)}
\label{Bint}%
\end{equation}
and
\begin{equation}
C=\int\limits_{\mathbf{q}_{1},\mathbf{q}_{2}}\frac{1}{\bigl(\mathbf{q}%
_{1}^{\,2}+1\bigr)^{2}\bigl(\mathbf{q}_{2}^{\,2}+1\bigr)^{2}\bigl(\mathbf{q}%
_{1}^{\,2}+(\mathbf{q}_{1}+\mathbf{q}_{2})^{2}+2\bigr)}\,. \label{Cint}%
\end{equation}
These integrals can be calculated for example by using Schwinger
parametrization, cf.\ the appendix. The $\varepsilon$ expansion results for
$B$ and $C$ read,
\begin{equation}
B=\frac{4G_{\varepsilon}^{2}}{9\varepsilon}\bigl(3\sqrt{3}-\pi\bigr)\,,\quad
C=\frac{4G_{\varepsilon}^{2}}{3\varepsilon}\,.
\end{equation}
The $I_{\cdots}$-contributions in Eq.~(\ref{dynSelf}) cancel and we finally
get
\begin{equation}
\label{diagramsRes}(\ref{DynGraph}a)+(\ref{DynGraph}b)=\frac{i\omega
G_{\varepsilon}^{2}g\tau^{-\varepsilon}}{3\varepsilon}\biggl(2-\frac{\sqrt{3}%
}{2}+\frac{\pi}{6}\biggr)\ .
\end{equation}

\subsection{Renormalization and renormalization group equation}

The dynamic theory requires an additional renormalization factor, say
$Z^{\prime}$, in comparison to the quasi-static theory due to the existence of
$\varepsilon$ poles in the frequency dependent terms. Taking into account that
the dynamic fields $n$ and $\widetilde{n}$ are related to the quasi-static
fields $s$ and $\widetilde{s}$ via Eq.~(\ref{quaslim}), we introduce
$Z^{\prime}$ consistent with the renormalization scheme in Eqs.~(\ref{RenFeld}%
) and (\ref{RenPar}) by letting
\begin{equation}
n\rightarrow{\mathring{n}}=Z^{\prime1/2}n\,,\quad\lambda\rightarrow
{\mathring{\lambda}}=(Z/Z^{\prime})^{1/2}\lambda\,. \label{RenDyn}%
\end{equation}
Combining Eq.~(\ref{diagramsRes}) with the zero-loop part of the vertex
function $\Gamma_{1,1}$ we get by applying our renormalizations
\begin{equation}
\Gamma_{1,1}=i\omega\biggl[\bigl(Z^{\prime}\widetilde{Z}\bigr)^{1/2}-\frac
{u}{6\varepsilon}\biggl(1+\frac{12-3\sqrt{3}}{\pi}\biggr)+O(u^{2})\biggr]\,.
\label{dynGam11}%
\end{equation}
To keep the formula~(\ref{dynGam11}) simple, we have once more  dropped all terms which are not linear in $\omega$. From Eq.~(\ref{dynGam11}) we can directly read off
$(Z^{\prime}\widetilde{Z})^{1/2}$ to order $u$. Taking into account
Eq.~(\ref{ZFakt}), this yields
\begin{equation}
Z^{\prime}=1+\frac{u}{\varepsilon}\biggl(\frac{4-\sqrt{3}}{\pi}-\frac{58}%
{15}\biggr)+O(u^{2})\ . \label{Zstrich}%
\end{equation}
The corresponding Wilson function reads
\begin{equation}
\gamma^{\prime}=2\biggl(\frac{58}{15}-\frac{4-\sqrt{3}}{\pi}\biggr)u+O(u^{2}%
)\,. \label{gamstr}%
\end{equation}
Now we have all the information required to calculate the Gell-Mann--Low
function
\begin{equation}
\zeta=\mu\left.  \partial_{\mu}\right\vert _{0}\ln\lambda=\frac{1}%
{2}\bigl(\gamma^{\prime}-\gamma\bigr) \label{zeta}%
\end{equation}
for the kinetic coefficient.

By proceeding analogous to the quasi-static case we obtain the dynamic RGE
\begin{equation}
\biggl[\mathcal{D}_{\mu}+\frac{1}{2}\bigl(N\gamma^{\prime}+\widetilde
{N}\widetilde{\gamma}\bigr)\biggr]G_{N,\widetilde{N}}(\{\mathbf{r}%
,t\},\tau)=0\,. \label{RGGdyn}%
\end{equation}
Here, the RG differential operator $\mathcal{D}_{\mu}$ is given by
\begin{equation}
\mathcal{D}_{\mu}=\mu\partial_{\mu}+\lambda\zeta\partial_{\lambda}%
+\bigl(\tau\kappa_{\tau}+v^{2}\kappa_{v\tau}\bigr)\partial_{\tau}+v\kappa
_{v}\partial_{v}+\beta_{u}\partial_{u}\,. \label{RGopdyn}%
\end{equation}

\subsection{Scaling properties}

\subsubsection{General scaling form}

Upon using the method of characteristics we obtain the dynamic scaling form
\begin{align}
&  G_{N,\widetilde{N}}(\{\mathbf{r},t\},\tau)=\ell^{\delta_{N,\widetilde{N}}%
}\mu^{(d-2)N+2\widetilde{N}}\nonumber\label{GreenDyn}\\
&  \times F_{N,\widetilde{N}}(\{\ell\mu\mathbf{r},\ell^{z}\lambda\mu
^{2}t\},\mu^{-2}\sigma/\ell^{1/\nu},\mu^{-1}v/\ell^{\phi/\nu})\,,
\end{align}
where
\begin{subequations}
\begin{align}
\delta_{N,\widetilde{N}}  &  =\Bigl(d-4+\frac{\eta}{2}+z\Bigr)N+\Bigl(2+\frac
{\widetilde{\eta}}{2}\Bigr)\widetilde{N}\nonumber\\
&  =\Bigl(\frac{\beta}{\nu}+z\Bigr)N+\frac{\beta^{\prime}}{\nu}\widetilde
{N}\,,\\
z  &  =2+\zeta^{\ast}=2-\biggl(\frac{8}{15}+\frac{4-\sqrt{3}}{\pi}%
\biggr)\frac{2\varepsilon}{45}+O(\varepsilon^{2})\nonumber\\
&  =2-0.0558\,\varepsilon+O(\varepsilon^{2})\,. \label{zet}%
\end{align}
The dynamic exponent $z$ is identical to the fractal dimension $D_{min}$ of
the minimal or chemical path, $D_{min}=z$~\cite{wirFraktaleDim}. Near the
tricritical point ($v=0$), we get from Eq.~(\ref{GreenDyn}) that the response and
correlation functions of the agent at time $t$ obey the scaling form
\end{subequations}
\begin{equation}
G_{N,\widetilde{N}}(\{\mathbf{r}\},t,\tau)=t^{-N(1+\beta/\nu_{s}%
)-\widetilde{N}\delta_{s}}f_{N,\widetilde{N}}(\{\mathbf{r}/t^{1/z}\},\tau
t^{1/\nu_{s}})\,, \label{DynKorr}%
\end{equation}
where
\begin{subequations}
\label{spreadExp}%
\begin{align}
\delta_{s}  &  =\frac{\beta^{\prime}}{\nu z}=1-\biggl(\frac{11}{3}%
-\frac{4-\sqrt{3}}{\pi}\biggr)\frac{\varepsilon}{45}+O(\varepsilon
^{2})\,,\label{delta-spr}\\
\nu_{s}  &  =\nu z=1+\biggl(\frac{5}{3}-\frac{4-\sqrt{3}}{\pi}\biggr)\frac
{\varepsilon}{45}+O(\varepsilon^{2})\,. \label{nu-spr}%
\end{align}

\subsubsection{Dynamic scaling behavior of various percolation observables}
\label{dynScalBelow5}

First we will consider the spreading of the agent emanating from a localized
seed at $\mathbf{r}=0$ and $t=0$. Later on, we will turn to the case that the
initial state at time $t=0$ is prepared with a homogeneous initial density
$\rho_{0} $.

The survival probability $\mathcal{P}(t,\tau)$ that a cluster grown from a single seed
is still active at time $t$ can be derived from the field theoretic
correlation functions by using~\cite{Ja03}
\end{subequations}
\begin{equation}
\mathcal{P}(t)=-\lim_{k\rightarrow\infty}\langle\mathrm{e}^{-kN(0)}\tilde
{n}(-t)\rangle\,,
\end{equation}
where now $N(0)=\int d^{d}x\,n(\mathbf{x},0)$. By proceeding analogous to the
static case, i.e., by incorporating the term $\exp(-kN(0))$ into the dynamic
functional, one obtains
\begin{align}
\mathcal{P}(t,\tau)  &  =-\lim_{k\rightarrow\infty}\langle\tilde{s}%
(-t)\rangle_{k}\nonumber\label{newSurvProb}\\
&  =-G_{0,1}(\mathbf{-}t,\tau,k=\infty,u;\lambda,\mu)\,,
\end{align}
where $\langle\cdots\rangle_{k}$ denotes averaging with respect to the new
dynamic functional $\mathcal{J}_{k}$ that has absorbed the source term
featuring $k$. Equation~(\ref{DynKorr}) then implies that the survival
probability obeys the scaling form
\begin{equation}
\mathcal{P}(t,\tau)=t^{-\delta_{s}}f_{\mathcal{P}}(\tau t^{1/\nu_{s}})\,.
\label{SurvProb}%
\end{equation}
In the percolating phase ($\tau<0$) the survival probability tends to the
percolation probability for large $t$, $\mathcal{P}(t,\tau)\rightarrow
\mathcal{P}_{\infty}(\tau)\sim\left\vert \tau\right\vert ^{\beta^{\prime}}$.
Hence, the universal scaling function $f_{\mathcal{P}}(x)$ behaves as
$x^{\beta^{\prime}}$ for large values of $x$.

The mean density of the agent at time $t$ grown from a single seed follows
from Eq.~(\ref{DynKorr}) with $N=\widetilde N = 1$ as
\begin{equation}
\rho(\mathbf{r},t,\tau)=t^{-1-(\beta+{\beta}^{\prime})/\nu z}f(r/t^{1/z},\tau
t^{1/\nu z})\,. \label{density}%
\end{equation}

Knowing the density of the agent, we have immediately access to the number of
infected or growth sites
\begin{equation}
\mathcal{N}(t,\tau)=\int d^{d}r\,\rho(\mathbf{r},t,\tau) \label{number1}%
\end{equation}
which can be viewed as the average size of the epidemic at time $t$.
Equation~(\ref{density}) implies that
\begin{equation}
\mathcal{N}(t,\tau)=t^{\eta_{s}}f_{\mathcal{N}}(\tau t^{1/\nu_{s}})\,,
\label{GrS}%
\end{equation}
where
\begin{equation}
\eta_{s}=\frac{\gamma}{\nu z}-1=\biggl(\frac{1}{3}+\frac{4-\sqrt{3}}{\pi
}\biggr)\frac{\varepsilon}{45}+O(\varepsilon^{2})\,. \label{eta-spr}%
\end{equation}
In the non-percolating phase ($\tau>0$), the integral $\int_{0}^{\infty}%
dt\,{}\mathcal{N}(t,\tau)$ is proportional to the mean mass $\langle S\rangle$
of the static clusters of the debris, Eq.~(\ref{MeanMass}). The mean mass of
the debris at time $t$, on the other hand, is given by
\begin{equation}
\mathcal{M}(t,\tau)=\int_{0}^{t}dt^{\prime}\,\mathcal{N}(t^{\prime},\tau).
\label{defConvInd}%
\end{equation}
For $\mathcal{M}$ we find the scaling form
\begin{equation}
\mathcal{M}(t,\tau)=t^{\bar{\eta}_{s}}f_{\mathcal{M}}(\tau t^{1/\nu_{s}})
\label{number2}%
\end{equation}
with
\begin{equation}
\bar{\eta}_{s}=1+{\eta}_{s}=\frac{\gamma}{\nu_{s}}=\frac{\gamma}{\nu z}\,.
\label{eta-spr2}%
\end{equation}
The scaling functions $f_{\mathcal{N}}(x)$ and $f_{\mathcal{M}}(x)$ are
regular for small $x$. For $x\rightarrow\infty$ we learn from $\lim
_{t\rightarrow\infty}\mathcal{M}(t,\tau)=\mathcal{M}(\tau)=\langle S\rangle$
in conjunction with Eq.~(\ref{MeanMass}) that
\begin{subequations}
\label{Skal-Fu}%
\begin{align}
f_{\mathcal{M}}(x)  &  =x^{-\gamma}f_{\mathcal{M}}^{+}(x^{\nu z}%
),\label{Skal-Fu1}\\
\mathcal{M}_{0}-f_{\mathcal{M}}^{+}(y)  &  \sim\exp(-\mathrm{const}\, y).
\label{Skal-Fu2}%
\end{align}
It follows that the number of growth sites behaves for $\tau>0$ asymptotically
as%
\end{subequations}
\begin{equation}
\mathcal{N}(t,\tau)\sim\tau^{\nu z-\gamma}\exp(-\mathrm{const} \, \tau^{\nu z}t).
\end{equation}

Knowing the density of the agent, we are in the position to calculate the mean
square distance $\mathcal{R}^{2}(t,\tau)$ of the infected individuals from the
original seed by using
\begin{equation}
\mathcal{R}^{2}(t,\tau)=\frac{1}{2d\, \mathcal{N}(t,\tau)}\int d^{d}%
r\,\mathbf{r}^{2}\rho(\mathbf{r},t,\tau)\,. \label{distance}%
\end{equation}
We obtain the scaling behavior
\begin{equation}
\mathcal{R}^{2}(t,\tau)=t^{z_{s}}f_{\mathcal{R}}(\tau t^{1/\nu_{s}})\,,
\end{equation}
where
\begin{equation}
z_{s}=\frac{2}{z}=1+\biggl(\frac{8}{15}+\frac{4-\sqrt{3}}{\pi}\biggr)\frac
{\varepsilon}{45}+O(\varepsilon^{2})\,.
\end{equation}

Mendes \emph{et al}.~\cite{MDHM94} proposed a generalized hyperscaling
relation which translates in our case to
\begin{equation}
2\biggl(1+\frac{\beta}{\beta^{\prime}}\biggr)\delta_{s}+2\bar{\eta}_{s}%
=dz_{s}\,. \label{Hyp}%
\end{equation}
From Eqs.~(\ref{MeanMass}), (\ref{SurvProb}), (\ref{number2}), and
(\ref{distance}) we we confirm that the spreading exponents indeed fulfill
this relation. This is not a surprise because the hyperscaling relation
(\ref{Hyp}) is based only on the general scaling form (\ref{DynKorr}).

Finally, we consider the scaling behavior of the time dependent mean density
of the agent $\rho(t,\tau,\rho_{0})=\langle n(\mathbf{r},t)\rangle_{\rho_{0}}$
for $t>0$ if the initial state at time $t=0$ is prepared with a homogeneous
initial density $\rho_{0}$. As mentioned earlier, this initial condition
corresponds to a source term $q(\mathbf{r},t)=\rho_{0}\delta(t)$ in the
Langevin equation~(\ref{eq:General Langevin}). This source term translates
into a further additive contribution $-\rho_{0}\int d^{d}r\,\widetilde
{n}(\mathbf{r},0)$ to the dynamic functional~(\ref{J}). In a theory like ours,
where the perturbation expansion is based only on causal propagators and where
no correlators appear, no initial time UV-infinities are generated. Therefore,
no independent short time scaling behavior \cite{Ja92,JSS89} arises and
$\widetilde{n}(\mathbf{r},0)$ scales as $\widetilde{n}(\mathbf{r},t)$. Thus,
we find, analogous to Eq.~(\ref{KorFuhom2}), that the dependence of the
correlation functions on $\rho_{0}$ can be expressed as
\begin{align}
&  C_{N}(\{\mathbf{r}\},t,\tau,\rho_{0})=t^{-(1+\beta/\nu z)N}\nonumber\\
&  \times f_{N}\big(\{\mathbf{r}/t^{1/z}\},\tau t^{1/\nu z},\rho_{0}%
t^{(d\nu-{\beta}^{\prime})/\nu z}\big)\,. \label{DynKorr2}%
\end{align}
In particular, we obtain for the mean density of the agent
\begin{equation}
\rho(t,\tau,\rho_{0})=t^{-1-\beta/\nu z}f_{\rho}\big(\tau t^{1/\nu z},\rho
_{0}t^{(d\nu-{\beta}^{\prime})/\nu z}\big)\,. \label{agdens}%
\end{equation}
At criticality ($\tau=0$) it follows from this equation that the agent density
first increases in the universal initial time regime,
\begin{equation}
\rho(t,\rho_{0})\sim\rho_{0}\,t^{\eta_{s}}\,. \label{initial}%
\end{equation}
Then, after some crossover time, it decreases,
\begin{equation}
\rho(t,\rho_{0})\sim t^{-1-\beta/\nu z}\,, \label{infscal}%
\end{equation}
with the critical exponent
\begin{equation}
1+\frac{\beta}{\nu z}=\frac{3}{2}-\biggl(\frac{107}{3}-\frac{4-\sqrt{3}}{\pi
}\biggr)\frac{\varepsilon}{90}+O(\varepsilon^{2})\,.
\end{equation}

The time dependence of the order parameter, the density of the converted
individuals
\begin{equation}
\bar{\rho}(t,\tau,\rho_{0})=\int_{0}^{t}dt^{\prime}\,\rho(t^{\prime},\tau
,\rho_{0})\,, \label{orddens1}%
\end{equation}
follows from Eq.~(\ref{agdens}),
\begin{equation}
\bar{\rho}(t,\tau,\rho_{0})=t^{-\beta/\nu z}f_{\bar{\rho}}\big(\tau t^{1/\nu
z},\rho_{0}t^{(d\nu-\widetilde{\beta})/\nu z}\big)\,. \label{orddens2}%
\end{equation}
This scaling form goes exponentially to the time independent scaling
form~(\ref{Dichte}) in the large time limit. Equation~(\ref{orddens2})
implies
\begin{equation}
\bar{\rho}(t,0,\rho_{0})\sim\rho_{0}\,t^{\theta}\,, \label{inord}%
\end{equation}
for the initial time order parameter scaling at criticality. The scaling index
appearing here is
\begin{equation}
\theta=\frac{\gamma}{\nu z}\,. \label{inscal}%
\end{equation}

\subsection{Logarithmic corrections in 5 dimensions}

Here we are going to investigate how logarithmic corrections influence the
dynamic scaling behavior in $d=5$. We will briefly explain how the general
considerations on logarithmic corrections in the quasi-static theory given in
Sec.~\ref{genFormStaic} have to be augmented and modified in the dynamic
theory. Then we will derive the logarithmic corrections for all of the dynamic
observables studied above.

\subsubsection{General form of the logarithmic corrections}

Compared to the quasi-static theory, $X^{\prime}$ takes on the role of $X$ and
there is an additional flowing variable, viz.\ $\lambda$. Both $X^{\prime}$
and $\lambda$ have characteristic equations of the form given in
Eq.~(\ref{Qstruct}) and hence the flow of these variables in $d=5$ is
described by Eq.~(\ref{Q(w)}). The dynamic scaling form~(\ref{GreenDyn})
becomes
\begin{align}
&  G_{N,\widetilde{N}}(\{\mathbf{r},t\},\tau)=(\mu\ell)^{3N+2\widetilde{N}%
}X^{\prime}(w)^{N/2}\widetilde{X}(w)^{\widetilde{N}/2}%
\nonumber\label{GreenDyn5}\\
&  \times F_{N,\widetilde{N}}\Big(\left\{  \mu\ell\mathbf{r},(\mu\ell
)^{2}\lambda(w)t\right\}  ,\frac{\tau(w)}{(\mu\ell)^{2}},w\Big).
\end{align}
in 5 dimensions. Since we are interested solely in the leading logarithmic corrections, it is sufficient for our purposes to account for the explicit dependence of the scaling functions $F_{N,\widetilde{N}}$ on $w$ to 0-Loop order, see Eq.~(\ref{Dangerous}). Using the asymptotic result (\ref{w-as}) we obtain%
\begin{align}
&G_{N,\widetilde{N}}(\left\{  \mathbf{r,t}\right\}  ,\tau)   =
\nonumber \\
& \ell
^{3N+2\tilde{N}}\left\vert \ln\ell\right\vert ^{7\tilde{N}%
/75+(a-22/225)N+(1-\tilde{N})/2}\nonumber\\
&  \times F_{N,\tilde{N}}\bigl(\{\ell\mathbf{r,\ell}^{2}\left\vert \ln
\ell\right\vert ^{a} t\},\ell^{-2}\left\vert \ln\ell\right\vert ^{11/225}%
\tau\bigr),\label{DynLogSkal}%
\end{align}
where $a$ is the abbreviation of%
\begin{equation}
a=\frac{1}{45}\Big(\frac{8}{15}+\frac{4-\sqrt{3}}{\pi}\Big)=0.0279.
\end{equation}

As it was in Sec.~\ref{dynScalBelow5}, our emphasize here is on the time
dependence of various measurable quantities. Hence we fix the arbitrary flow
parameter $\ell$ as long as $\lambda\left\vert \tau\right\vert t\leq 1$ by
setting
\begin{equation}
\ell^{2}\sim t^{-1}\,(\ln t)^{-a}. \label{choiceT}%
\end{equation}
In the contrary case $\lambda\left\vert \tau\right\vert t\gg1$ we must use the
choice (\ref{lTau}).

\subsubsection{Logarithmic corrections to dynamic percolation observables}

\label{dynIn5} As above we will first consider the initial condition that the
process starts from a single local seed at the space-wise and time-wise
origin. The first quantity that we are going to consider is the survival
probability $\mathcal{P}(t,\tau)$. Utilizing Eq.~(\ref{newSurvProb}) in
conjunction with Eq.~(\ref{GreenDyn5}) we obtain
\begin{equation}
\mathcal{P}(t,\tau)=\ell^{2}\left\vert \ln\ell\right\vert ^{7/75}%
Y_{\mathcal{P}}\bigl(\ell^{2}\left\vert \ln\ell\right\vert ^{a}t,\ell
^{-2}\left\vert \ln\ell\right\vert ^{-11/225}\tau\bigr).
\end{equation}
The choice~(\ref{choiceT}) then leads to
\begin{equation}
\mathcal{P}(t,\tau)=t^{-1}\,(\ln t)^{a_{\mathcal{P}}}\,f_{\mathcal{P}}\bigl(\tau
\,t\,(\ln t)^{a_{\tau}}\bigr),
\end{equation}
with the exponents
\begin{subequations}
\begin{align}
a_{\mathcal{P}} &  =\frac{7}{75}-a=\frac{1}{45}\Big(\frac{11}{3}-\frac
{4-\sqrt{3}}{\pi}\Big),\\
a_{\tau} &  =a-\frac{11}{225}=\frac{1}{45}\Big(-\frac{5}{3}+\frac{4-\sqrt{3}%
}{\pi}\Big).
\end{align}
Hence, we have at criticality and near criticality with $\lambda\left\vert
\tau\right\vert t\ll1$%
\end{subequations}
\begin{equation}
\mathcal{P}(t,\tau\approx0)\sim t^{-1}\,(\ln t)^{a_{\mathcal{P}}}.
\end{equation}
The asymptotic behavior for large times below and above criticality,
respectively, is given by%
\begin{subequations}
\begin{align}
\mathcal{P}(t,\tau &  >0)\sim\tau\left\vert \ln\tau\right\vert ^{2/45}%
\,\exp\bigl(-\mathrm{const}\, t\, \tau\left\vert \ln\tau\right\vert ^{a_{\tau}%
}\bigr)
\end{align}
and
\begin{align}
\mathcal{P}(t,\tau &  <0)-\mathcal{P}_{\infty}(-\left\vert \tau\right\vert
)\nonumber\\
&  \sim\left\vert \tau\right\vert \left\vert \ln\left\vert \tau\right\vert
\right\vert ^{2/45}\,\exp\bigl(-\mathrm{const}\, t\left\vert \tau\right\vert
\left\vert \ln\left\vert \tau\right\vert \right\vert ^{a_{\tau}}\bigr).
\end{align}

Next we look at the mean density of the active particles at time $t$. Upon
specializing the general scaling form~(\ref{DynLogSkal}) to $N=\widetilde
{N}=1$ we find
\end{subequations}
\begin{equation}
\rho(\mathbf{r},t,\tau)=\ell^{5}\left\vert \ln\ell\right\vert ^{a_{\mathcal{N}%
}}f_{\rho}\bigl(\ell\mathbf{r},\ell^{2}\left\vert \ln\ell\right\vert
^{a}t,\ell^{-2}\left\vert \ln\ell\right\vert ^{-11/225}\tau\bigr)
\end{equation}
with%
\begin{equation}
a_{\mathcal{N}}=a-\frac{1}{225}=\frac{1}{45}\Big(\frac{1}{3}+\frac{4-\sqrt{3}%
}{\pi}\Big)
\end{equation}
From the mean density we obtain the mean number of agents (\ref{number1}) at
time $t$ by integrating over $\mathbf{r}$, and choosing (\ref{choiceT}) for
$\tau\approx0$%
\begin{equation}
\mathcal{N}(t,\tau)=\left\vert \ln t\right\vert ^{a_{\mathcal{N}}%
}f_{\mathcal{N}}\bigl(\tau\,t\,(\ln t)^{a_{\tau}}\bigr)\,.
\end{equation}
For the mean mass of the debris at time $t$ as defined in
Eq.~(\ref{defConvInd}) we get
\begin{equation}
\mathcal{M}(t,\tau)=t\left\vert \ln t\right\vert ^{a_{\mathcal{N}}%
}f_{\mathcal{M}}\bigl(\tau\,t\,(\ln t)^{a_{\tau}}\bigr)\,.
\end{equation}
$\mathcal{M}(t,\tau)$ crosses over for $\tau>0$ and $t\rightarrow\infty$ to a
function which approaches $\mathcal{M}(\tau)=\langle S\rangle$, Eq.~(\ref{meanClM}), exponentially.
The mean square distance ${\mathcal{R}}^{2}(t,\tau)$ of the agents from the
origin is found to behave as
\begin{equation}
\mathcal{R}^{2}(t,\tau)=t\,(\ln t)^{a}\,f_{\mathcal{R}}\bigl(\tau\,t\,(\ln
t)^{a_{\tau}}\bigr).
\end{equation}

Finally, we will consider the homogeneous initial condition. The analog of
Eq.~(\ref{DynKorr2}) in 5 dimensions with the flow parameter $\ell$ still
arbitrary reads 
\begin{align}
&  C_{N}(\{\mathbf{r}\},t,\tau,\rho_{0})=\ell^{3N}\left\vert \ln
\ell\right\vert ^{(a-22/225)N+1/2}\nonumber\label{DynKorr25}\\
&  \times F_{N}\bigl(\{\ell\mathbf{r,\ell}^{2}\left\vert \ln\ell\right\vert
^{a}t\},\ell^{-2}\left\vert \ln\ell\right\vert ^{-11/225}\tau,\ell
^{-3}\left\vert \ln\ell\right\vert ^{-61/150}\rho_{0}\bigr)
\end{align}
where we have used Eq.~(\ref{DynLogSkal}). From Eq.~(\ref{DynKorr25}) we readily obtain the critical behavior of the mean
density of the agents by setting $N=1$ and fixing $\ell$ for not too large $t$
via Eq.~(\ref{choiceT}),
\begin{equation}
\rho(t,\tau,\rho_{0})=t^{-3/2}\,(\ln t)^{a_{\rho}}\,f_{\rho}\bigl(\tau
\,t\,\,(\ln t)^{a_{\tau}},\rho_{0}\,t^{3/2}\,(\ln t)^{-a_{\rho}^{0}%
}\bigr),\label{densGen}%
\end{equation}
where
\begin{subequations}
\begin{align}
a_{\rho} &  =\frac{181}{450}-\frac{a}{2}=\frac{1}{90}\Big(\frac{107}{3}%
-\frac{4-\sqrt{3}}{\pi}\Big),\\
a_{\rho}^{0} &  =\frac{61}{150}-\frac{3a}{2}=\frac{1}{30}\Big(\frac{35}%
{3}-\,\frac{4-\sqrt{3}}{\pi}\Big).
\end{align}
At criticality, the scaling function is expected to behave as $f_{\rho
}(0,y)\sim y$ for $y\ll1$. Thus, the agent density increases initially,
\end{subequations}
\begin{equation}
\rho(t,0,\rho_{0})\sim\rho_{0}\,(\ln t)^{a_{\mathcal{N}}}.
\end{equation}
After some crossover time it decreases like
\begin{equation}
\rho(t,0,\rho_{0})\sim t^{-3/2}\,(\ln t)^{a_{\rho}}\,.
\end{equation}
The mean density of the debris at time $t$ can be extracted without much
effort by integrating over Eq.~(\ref{densGen}). This yields
\begin{equation}
\bar{\rho}(t,\tau,\rho_{0})=t^{-1/2}\,(\ln t)^{a_{\rho}}\,\bar{f}_{\rho
}\bigl(\tau\,t\,\,(\ln t)^{a_{\tau}},\rho_{0}\,t^{3/2}\,(\ln t)^{-a_{\rho}%
^{0}}\bigr)\label{densGenConv}%
\end{equation}
if $t$ is not too large. In the case $\rho_{0}\,t^{3/2}\rightarrow\infty$ we
have to use
\begin{equation}
\ell^{3}\sim\rho_{0}\left\vert \ln\rho_{0}\right\vert ^{-61/150},
\end{equation}
which implies especially at criticality%
\begin{equation}
\bar{\rho}(t,\tau=0,\rho_{0})=\rho_{0}^{1/3}\left\vert \ln\rho_{0}\right\vert
^{4/15}\,\bar{f}_{\rho}^{\prime}\bigl(t\rho_{0}^{2/3}\,\left\vert \ln\rho
_{0}\right\vert ^{-2a_{\rho}^{0}/3}\bigr).
\end{equation}
The scaling function $\bar{f}_{\rho}^{\prime}(x)$ approaches exponentially the
stationary density (\ref{StatDens}).

\section{Concluding Remarks}

\label{sec:conclusion}

In summary, we have generalized the usual GEP by introducing a further state
in the live of the individuals governed by the process. Our GGEP has a
multi-dimensional phase diagram featuring two surfaces separating endemic and
pandemic behavior of the epidemic. One of the surfaces is a surface of first
order phase transitions whereas the other surface consists of critical points
representing second order transitions. The two surfaces meet at a line of
tricritical points.

The second order phase transitions belong to the universality class of dynamic
isotropic percolation (dIP). In the vicinity of these transitions, the
asymptotic time limit of the GGEP is governed by the critical exponents of
usual percolation. The debris left behind by the process forms isotropic
percolation clusters.

Mainly, we were interested in the tricritical behavior of the GGEP. We set up
a field theoretic minimal model in the form of a dynamic response functional
that allowed us to study in detail the static and the dynamic scaling behavior
of the universality class of tricritical dynamic isotropic percolation (TdIP). In particular we calculated the
scaling exponents for various quantities that play an important role in
percolation theory. As expected, these exponents are different from the
exponents pertaining to dIP. For example, we computed the
exponents $\beta$ and $\beta^{\prime}$ respectively describing the two
different order parameters, viz.\ the density of the debris and the
percolation probability. Whereas $\beta$ and $\beta^{\prime}$ are identical in dIP, they are different in TdIP. Although TdIP is described by scaling exponents different from those of usual
percolation, we learned that its spreading as well as the statistics of its
clusters behave in many ways like conventional dynamic percolation. For
example, TdIP has meaningful cluster numbers, fractal dimensions etc.\ Thus,
we propose to refer to the static properties of the TdIP as tricritical
isotropic percolation (TIP).

The surface of first order transitions is characterized by a compact cluster
growth, i.e., the fractal dimension of the clusters is identical to the their
embedding dimension. We hope that our findings trigger numerical work with the
aim to verify the predicted first order percolation transitions. A promising
strategy that avoids a cumbersome detection of jumps in the order parameters
might be to measure directly the fractal dimension of clusters near the first
order surface.

\begin{acknowledgments}
This work has been supported by the Deut\-sche For\-schungs\-ge\-mein\-schaft
via the Son\-der\-for\-schungs\-be\-reich~237 ``Un\-ord\-nung und gro\-{\ss }e
Fluk\-tu\-a\-ti\-o\-nen''.
\end{acknowledgments}

\appendix*

\section{Calculation of the parameter integral $I(a, b; c)$}

In this appendix we sketch our calculation of the parameter integral $I(a, b;
c)$ defined in Eq.~(\ref{parameterI}). Most of the integrals that have to be
performed in calculating the two-loop diagrams can be derived from $I(a, b;
c)$ simply by taking derivatives with respect to the parameters $a$, $b$, and
$c$. The integrals $B$ and $C$, which occur in the dynamic calculation and
cannot be extracted from $I(a, b; c)$ by taking derivatives, can be calculated
by similar means as $I(a, b; c)$.

In the following we use the so-called Schwinger parametrization which is based
on the identity
\begin{align}
\frac{1}{A^{n}} = \frac{1}{\Gamma(n)} \, \int_{0}^{\infty}d s\, t^{n-1}
\exp(-As)\, , \quad\mbox{Re} \, n > 0 \, .
\end{align}
In this parametrization, Eq.~(\ref{parameterI}) takes on the form
\begin{align}
I(a,b;c)  &  =\int\limits_{\mathbf{q}_{1},\mathbf{q}_{2}} \, \int_{0}^{\infty
}d s_{1}d s_{2} d s_{3} \, s_{3} \exp\big\{ - s_{1} [\mathbf{q}_{1}^{2} +
a]\nonumber\\
&  - s_{2} [\mathbf{q}_{2}^{2} + b] - s_{1} [ (\mathbf{q}_{1} + \mathbf{q}_{2}
)^{2} + c]\big\} \, .
\end{align}
A completion of squares in the momenta renders the momentum integrations
straightforward. We obtain
\begin{align}
I(a,b;c)  &  =\frac{1}{(4\pi)^{d}} \int_{0}^{\infty}d s_{1}d s_{2} d s_{3} \,
s_{3} \,\nonumber\\
&  \times\, \frac{\exp(-s_{1} a -s_{2} b -s_{3} c)}{[s_{1} s_{3} + s_{2} s_{3}
+ s_{1} s_{2}]^{d/2}} \, .
\end{align}
Changing integration variables, $s_{1} \to t x$, $s_{2} \to t (1-x)$ and
$s_{3} \to t z$, and carrying out the $t$-integration gives
\begin{align}
I(a,b;c)  &  =\frac{\Gamma(4-d)}{(4\pi)^{d}} \int_{0}^{\infty}d z \int_{0}^{1}
dx \, \frac{z}{[z + x(1-x)]^{d/2}}\nonumber\\
&  \times\, \big[ a x + b (1-x) + c z \big]^{d-4} \, .
\end{align}
The remaining integrations can be simplified by letting $z \to(z^{-1}
-1)x(1-x)$. After this step, which leads to
\begin{align}
I(a,b;c)  &  =\frac{\Gamma(4-d)}{(4\pi)^{d}} \int_{0}^{\infty}d z \int_{0}^{1}
dx\nonumber\\
&  \times\, x^{2-d/2} (1-x)^{2-d/2} z^{1-d/2} (1-z)\nonumber\\
&  \times\, \big[ a x z+ b (1-x) z+ c x (1-x) (1-z) \big]^{d-4} \, ,
\end{align}
one sees easily that the remaining integrations are finite at the upper
critical dimension. Hence, they can be conveniently evaluated directly at
$d=5$. An $\varepsilon$ expansion of the gamma function,
\begin{align}
\Gamma(4-d) = - \, \frac{\Gamma(1 + \varepsilon/2)^{2}}{\varepsilon} + O
(\varepsilon^{0})\, ,
\end{align}
finally leads to the result for $I(a,b;c)$ stated in Eq.~(\ref{mother}).


\begin{thebibliography}{99}                                                                                               %


\bibitem {BrHa57}S.R.\ Broadbent and J.M.\ Hammersley,
Proc.\ Camb.\ Philos.\ Soc.\ \textbf{53}, 629 (1957).

\bibitem {CaSu80}J.L.\ Cardy and R.L.\ Sugar, J.\ Phys.\ A:
Math.\ Gen.\ \textbf{13}, L423 (1980).

\bibitem {Ob80}S.P.\ Obukhov, Physica A \textbf{101}, 145 (1980).

\bibitem {Hin00}A comprehensive recent overview over directed percolation can
be found in H.\ Hinrichsen, Adv.\ Phys.\ \textbf{49}, 815 (2001).

\bibitem {GraSu78}P.\ Grassberger and K.\ Sundermeyer, Phys.\ Lett.\ B
\textbf{77}, 220 (1978).

\bibitem {GraTo79}P.\ Grassberger and A.\ De~La~Torre, Ann.\ Phys.\ (New York)
\textbf{122}, 373 (1979).

\bibitem {Schl72}F.\ Schl\"{o}gl, Z.\ Phys.\ \textbf{225}, 147 (1972).

\bibitem {Ja81}H.K.\ Janssen, Z.\ Phys.\ B: Cond.\ Mat.\ \textbf{42}, 151 (1981).

\bibitem {Gri67}V.N.\ Gribov, Zh.\ Eksp.\ Teor.\ Fiz.\ \textbf{53}, 654 (1967)
[Sov.\ Phys.\ JETP \textbf{26}, 414 (1968)].

\bibitem {GriMi68}V.N.\ Gribov and A.A.\ Migdal,
Zh.\ Eksp.\ Teor.\ Fiz.\ \textbf{55}, 1498 (1968) [Sov.\ Phys.\ JETP
\textbf{28}, 784 (1969)].

\bibitem {Mo78}M.\ Moshe, Phys.\ Rep.\ C\textbf{\ 37}, 255 (1978).

\bibitem {Mol77}D.\ Mollison, J.R.\ Stat.\ Soc.\ B \textbf{39}, 283 (1977).

\bibitem {Bai75}N.T.J.\ Bailey, \emph{The Mathematical Theory of Infectious
Diseases}, (Griffin, London, 1975)

\bibitem {Mur89}J.D.\ Murray, \emph{Mathematical Biology}, (Springer, Berlin 1989).

\bibitem {Gra83}P.\ Grassberger, Math.\ Biosci.\ \textbf{63}, 157 (1983).

\bibitem {Ja85}H.K. Janssen, Z.\ Phys.\ B: Cond.\ Mat.\ \textbf{58}, 311 (1985).

\bibitem {CaGra85}J.L.\ Cardy and P.\ Grassberger, J.\ Phys.\ A:
Math.\ Gen.\ \textbf{18}, L267 (1985).

\bibitem {StAh92BuHa96}For an introduction to and an overview over isotropic
percolation see, e.g., D.~Stauffer and A.~Aharony, \emph{Introduction to
Percolation Theory}, 2nd edition, (Taylor and Francis, London); A.\ Bunde and
S.\ Havlin, in \emph{Fractals and Disordered Systems}, 2nd edition,
eds.\ A.\ Bunde, S.\ Havlin (Springer, Berlin, 1996).

\bibitem{footnote_quenchedDisorder}
Regarding the SEP, quenched disorder of the substrate, i.e., frozen in local deviations of the reaction rates from their spatial average, perturbs the process crucially so that it no longer belongs to the DP universality class, see H.K. Janssen,  Phys.~Rev.~E \textbf{55}, 6253 (1997). With respect to dIP, on the other hand, quenched disorder of the substrate is irrelevant.

\bibitem {OhKe87}T.~Ohtsuki and T.~Keyes, Phys.~Rev.~A \textbf{35}, 2697
(1987); Phys.~Rev.~A \textbf{36}, 4434 (1987).

\bibitem{footnote_diff_sick}Wether the sick individuals diffuse or not does not alter the leading critical behavior of the GEP because the process spreads diffusively anyway via the infection of neighbors.

\bibitem{tomsBook}
See, e. g., P. M. Chaikin and T. C. Lubensky, {\em
Principles of Condensed Matter Physics}, (Cambridge
University Press, Cambridge, 1995).

\bibitem {Ed81}M.~Eden, in \emph{Proceedimgs of theFourth Berkeley Symposium
on Mathematical Statistics and Probability}, Vol.~IV, edited by J.~Neyman,
(University of California Press, Berkeley, 1981).

\bibitem {Ca83}J.L.\ Cardy, J.\ Phys.\ A: Math.\ Gen.\ \textbf{16}, L709 (1983).

\bibitem {Ja76}H.K.~Janssen, Z.~Phys.~B: Cond.\ Mat. \textbf{23}, 377 (1976);
R.~Bausch, H.K.~Janssen, and H.~Wagner, Z.~Phys.~B: Cond.\ Mat. \textbf{24},
113 (1976); H.K.~Janssen, in \emph{Dynamical Critical Phenomena and Related
Topics} (\emph{Lecture Notes in Physics}, Vol.~104), edited by C.P.~Enz,
(Springer, Heidelberg, 1979).

\bibitem {DeDo76}C.~De~Dominicis, J.~Phys.~(France) Colloq.~\textbf{37}, C247
(1976); C.~De~Dominicis and L.~Peliti, Phys.~Rev.~B \textbf{18}, 353 (1978).

\bibitem {Ja92}H.K.~Janssen, in \emph{From Phase Transitions to Chaos}, edited
by G.~Gy\"{o}rgyi, I.~Kondor, L.~Sasv\'{a}ri, and T.~T\'{e}l, (World
Scientific, Singapore, 1992).

\bibitem {Wils75}K.G.~Wilson, Rev.~Mod.~Phys.~\textbf{47}, 773 (1975).

\bibitem {Am84}D.J.~Amit, \emph{Field Theory, the Renormalization Group and
Critical Phenomena}, (World Scientific, Singapore, 1984).

\bibitem {ZiJu93}J.~Zinn-Justin, \emph{Quantum Field Theory and Critical
Phenomena}, 2nd revised edition, (Clarendon, Oxford, 1993).

\bibitem {JSS89}H.K.~Janssen, B.~Schaub, and B.~Schmittmann, Z.~Phys.~B:
Cond.\ Mat. \textbf{73}, 539 (1989).

\bibitem {JaSt04}H.K.~Janssen and O.~Stenull, Phys.~Rev.~E \textbf{69}, 016125 (2004).

\bibitem {wirFraktaleDim}For a discussion of chemical path in the field theory
of conventional isotropic percolation, see e.g., H.K.~Janssen, O.~Stenull, and
K.~Oerding, Phys.~Rev.~E \textbf{59}, R6239 (1999); H.K.~Janssen and
O.~Stenull, Phys.~Rev.~E \textbf{61}, 4821 (2000).

\bibitem {Ja03}H.K.~Janssen, cond-mat/0304631.

\bibitem {MDHM94}J.F.F.~Mendes, R.~Dickman, M.~Henkel, and M.C.~Marques,
J.\ Phys.\ A: Math.\ Gen.\ \textbf{27}, 3019 (1994).
\end{thebibliography}
\end{document}